\documentclass[referee]{aa} 
\newcommand\etal{{et al.~}}
\newcommand\asca{{\it ASCA~}}

\newcommand\chandra{{\it Chandra~}}

\newcommand\rosat{{\it ROSAT~}}

\newcommand\s{{\rm~s~}}
\newcommand\kev{{\rm~keV}}
\newcommand\ev{{\rm~eV}}
\newcommand\kms{\ifmmode {\rm~km\ s}^{-1} \else ~km s$^{-1}$\fi}
\newcommand\Hunit{\ifmmode {\rm~km\ s}^{-1}\ {\rm Mpc}^{-1}
        \else ~km s$^{-1}$ Mpc$^{-1}$\fi}
\newcommand\ctssec{\ifmmode {\rm~count\ s}^{-1} \else ~count s$^{-1}$\fi}
\newcommand\ergsec{\ifmmode {\rm~erg\ s}^{-1} \else
        ~erg s$^{-1}$\fi}
\newcommand\funit{\ifmmode {\rm~erg\ s}^{-1}\;{\rm cm}^{-2} \else
        ~ergs s$^{-1}$ cm$^{-2}$\fi}
\newcommand\phflux{\ifmmode {\rm~photon\ s}^{-1}\;{\rm cm}^{-2}
        \else   ~photon s$^{-1}$ cm$^{-2}$\fi}
\newcommand\efluxA{\ifmmode {\rm~erg\ s}^{-1}\;{\rm cm}^{-2}\;{\rm
        \AA}^{-1} \else ~erg s$^{-1}$ cm$^{-2}$ \AA$^{-1}$\fi}
\newcommand\efluxHz{\ifmmode {\rm~erg\ s}^{-1}\;{\rm cm}^{-2}\;{\rm
        Hz}^{-1} \else ~erg s$^{-1}$ cm$^{-2}$ Hz$^{-1}$\fi}
\newcommand\cc{\ifmmode {\rm~cm}^{-3} \else cm$^{-3}$\fi}
\newcommand\FWHM{\ifmmode {\rm~FWHM} \else ${\rm~FWHM}$\fi}
\newcommand\Msun{\ifmmode M_{\odot} \else $M_{\odot}$\fi}
\newcommand\Lsun{\ifmmode L_{\odot} \else $L_{\odot}$\fi}

\newcommand\hbeta{\ifmmode {\rm H}\beta \else H$\beta$\fi}
\newcommand\Kalpha{\ifmmode {\rm K}\alpha \else K$\alpha$\fi}
\newcommand\NH{\ifmmode N_{\rm H} \else N$_{\rm H}$\fi}

\newcommand\araa{ARA\&A}%
\newcommand\apj{ApJ}%
\newcommand\apjs{ApJS}%
\newcommand\aap{A\&A}%
\newcommand\mnras{MNRAS}%
\newcommand\pasj{PASJ}%
\usepackage{graphicx}
\begin{document}
   \title{A 10-day \asca Observation of the Narrow-line Seyfert~1 galaxy IRAS~13224-3809}

   \author{G. C. Dewangan 
          \inst{1}
          \and
          Th. Boller\inst{2}
	  \and
	  K. P. Singh\inst{1,4}
	  \and
	  K. M. Leighly\inst{3}
          }

   \offprints{G. C. Dewangan}

   \institute{Department of Astronomy \& Astrophysics, Tata Institute of Fundamental Research, Mumbai 400~005 India\\
              \email{gulab@tifr.res.in}
         \and
	   Max-Planck-Institute for Extraterrestrical Physics, Giessenbachstr.,
	   85748 Garching, Germany\\
             \email{bol@mpe.mpg.de}
	  \and
	   Department of Physics and Astronomy, The University
	   of Oklahoma, 440 W.\ Brooks St., Norman, OK 73072, USA\\
	     \email{leighly@ou.edu}
	  \and
	     \email{singh@tifr.res.in}
             }

   \date{Received ; accepted }

   \abstract{
   We present an analysis of a 10-day continuous \asca observation
   of the narrow-line Seyfert~1 galaxy IRAS~13224-3809. The
   total band ($0.7-10{\rm~keV}$) light curve binned with 500~s
   reveals trough-to-peak variation by a factor $\ge37$. Rapid X-ray
   variability with a doubling timescale of 500~s has also been detected.
   The soft ($0.7-1.3{\rm~keV}$) and hard ($1.3-10{\rm~keV}$) X-ray band
   light curves binned to 5000~s  reveal trough-to-peak variations
   by a factor $\ge 25$ and $\sim20$, respectively. The light
   curves in the soft and hard bands are strongly correlated without any significant delay. However, this
   correlation is not entirely due to changes in the power-law flux alone but also
   due to changes in the soft X-ray hump emission above the power law.
   The variability
   amplitude changes across the observation but is not correlated with
    the X-ray flux. The presence of a soft X-ray hump below $\sim2{\rm~keV}$, previously
    detected in \rosat and \asca data, is confirmed. Time resolved
    spectroscopy using daily sampling reveals changes in the power-law slope,
    with $\Gamma_{\rm X}$ in the range $1.74-2.47$, however, day-to-day variations
    in $\Gamma_{\rm X}$ are not significant. 
   The Soft hump emission is found to dominate the observed variability on a timescale of $\sim a~week$, but on shorter timescales ($\sim 20000\s$) the power-law component appears to dominate the observed variability.
   Flux resolved
    spectroscopy reveals that at high flux levels the power law becomes steeper
    and the soft hump more pronounced. This result is further confirmed using
    an earlier \asca observation in 1994. The   
     steepening of the photon-index with the  
     fluxes in the soft and hard bands can be understood in the framework of disk/corona models in which
    accretion disk is heated by viscous dissipation as well as by reprocessing 
    of hard X-rays following an X-ray flare resulting from coronal dissipation 
    through magnetic reconnection events. Time dependent accretion disk-corona models are required to understand the observed correlation between the soft hump emission and the power-law flux.
   }

   \maketitle
%

\section{Introduction}

Seyfert~1 galaxies are an important class of active galactic nuclei (AGN). 
They show a large range in the width of their optical emission lines e.g., full 
width at half maximum ($\FWHM$) of the $\hbeta$ line is found to be 
in the range $\sim1000-10000{\rm~km~s^{-1}}$. Seyfert~1 galaxies that are
at the lower end of the line width distribution with 
$\FWHM(\hbeta) \la2000{\rm~km~s^{-1}}$ are called the narrow-line
Seyfert~1 galaxies (NLS1; Osterbrock \& Pogge 1985; Goodrich 1989) and are
distinguished from the bulk of the Seyfert~1s (``broad-line Seyfert~1s''
or BLS1s). The NLS1 galaxies are also found to have strong Fe~II emission 
and  [O~III]$\lambda5007$/H$\beta < 3$ (Osterbrock \& Pogge 1985; 
Goodrich 1989). However, Rodriguez-Ardilla \etal (2000) have shown that
the ratio [O~III]$\lambda5007$/H$\beta$ does not clearly distinguish 
between NLS1 and BLS1. Also Veron-Cetty, Veron \& Goncalves (2001) have shown
that when only the narrow component of H$\beta$ is considered, the above ratio
is similar between NLS1s and Seyfert~2 galaxies. X-ray properties of NLS1s 
are even more remarkable. These AGNs very frequently exhibit rapid and/or 
large amplitude variability (Boller \etal 1996; 
Forster \& Halpern 1996; Molthagen \etal 1998). The excess 
variance for NLS1s is typically an order of magnitude higher than that observed
for samples of BLS1s with similar luminosity distribution (Leighly 1999a; Turner \etal 1999b). Giant-amplitude X-ray variability (up to a factor of 100)
has also been observed in several NLS1s (Boller \etal 1997; Brandt \etal 1999).
Some NLS1 galaxies show extremely rapid variability (on timescales of a few hundred seconds) by a factor of about 2--3 (Remillard \etal 1991; Boller \etal 1997; Brandt \etal 1999; Dewangan \etal 2001a). \rosat ($0.1-2.4\kev$) observations
have revealed that the soft X-ray continuum slopes of NLS1s are systematically 
steeper than those of BLS1s (Boller \etal 1996), the photon index $\Gamma_{\rm X}$ (photon flux $f_E \propto E^{- \Gamma_{\rm X}}$) sometimes exceeding 3. \asca 
observations have shown that the hard X-ray $2-10\kev$ continuum slope too is 
significantly steeper in NLS1s than that in the BLS1s (Brandt \etal 1997; Turner \etal 1998; Leighly 1999b; Vaughan \etal 1999b). The
very strong anti-correlation between $\FWHM$ of the H$\beta$ line and both the X-ray slopes in Seyfert~1s (Boller \etal 1996) and in quasars (Laor \etal 1997), and ``excess variance'' (Turner \etal 1999b) suggests that the remarkable X-ray properties of NLS1s are possibly due to an extreme value of a fundamental physical parameter related to the accretion process.

A popular explanation for the distinct properties of NLS1 galaxies is that they have lower black-hole masses than the BLS1 galaxies. Smaller black-hole masses result in shorter timescales, thus naturally explaining the rapid X-ray variability, since the primary emission would originate in a smaller region around the central black-hole. Smaller black-hole masses also naturally result in narrower optical emission lines provided the size of the broad emission line region (BLR) scales with the luminosity (Laor 1998).

A comparison of the soft X-ray properties of Seyferts and Galactic 
black-hole candidates (GBHCs) led Pounds \etal (1995) 
to make an analogy between the two types of objects. They suggested 
that NLS1s are the high state analog of BLS1 galaxies. The high state 
GBHCs show strong soft X-ray excess, with blackbody temperature 
$\sim2\kev$, above a steep power law and are thought to emit a higher 
fraction of their Eddington luminosity. This led Pounds \etal (1995) 
to postulate that NLS1s must also be emitting a higher fraction of 
their Eddington luminosity, hence higher accretion rates relative to the Eddington accretion rate ($\dot{m} = \frac{\dot{M}}{{\dot{M}_{\rm Edd}}}$) are required. Since NLS1s have comparable luminosity to that of BLS1s, a higher fractional rate also means a lower black-hole mass.

The higher the fraction of the Eddington luminosity emitted, i.e. the higher the fractional accretion rate, the greater the temperature attained by the accretion disk, i.e. the disk emission becomes energetically dominant in the soft X-rays (Ross \etal 1992). Thus NLS1s might be expected to show disk components which peak at higher energies than for BLS1s. The spectral energy distribution (SED) from far-infrared to X-rays of NLS1 galaxies appears to be similar to that of BLS1 galaxies, but the UV luminosity of NLS1s tends to be smaller than that of BLS1s (Rodriguez-Pascual \etal 1997). The lower UV luminosity
of NLS1 galaxies compared to that of BLS1s could be due to the shift of the big blue bump (BBB) towards higher energies. The steep soft X-ray spectrum could be the high energy tail of the BBB (Mathur 2000). Pounds \etal (1995) noted that the excess soft X-ray emission of NLS1s may cause an increased Compton cooling of hot electrons in the corona resulting in a steeper hard X-ray power law. Higher accretion rates also result in an ionized surface for the accretion disk (Matt \etal 1993). Evidence for the ionized disk is found in the form of K$\alpha$ emission from the ionized states of Fe in NLS1s (Comastri \etal 1998; Turner \etal 1998; Vaughan \etal 1999a; Comastri \etal 2001; Turner \etal 1999a; Ballantyne \etal 2001). However, the ionized Fe~K$\alpha$ line is not unique to NLS1 galaxies, and some BLS1 galaxies also show ionized Fe~K$\alpha$. This may suggest that the luminosity of the central source plays as important a role as the accretion rate (e.g., Guainazzi \etal 1998).

Alternative explanations for the extreme properties of NLS1s are: 
($i$) the size of the BLR of NLS1s is larger (i.e., the BLR gas is 
more distant from the nucleus) than that in the BLS1s 
(Guilbert, Fabian, \& McCray 1983; Mason \etal 1996; 
Wandel \& Boller 1998) resulting in the narrowness of the width of the 
permitted lines due to a lower orbital velocity; ($ii$) we have a nearly 
face-on view of a flattened BLR in NLS1s (Osterbrock \& Pogge 1985). 
Assuming the motion of the BLR gas around the central super-massive black-hole 
to be virialized, the narrowness of the lines is due to the fact that the 
gas is moving preferentially on a plane that is almost perpendicular 
to the line of sight resulting in the smaller velocity dispersions, hence 
the line widths are reduced by a factor sin$i$, where $i=0$ is face-on.
However, Boroson \& Green (1992) and Kuraszkiewcz \etal (2000) do not favor the low inclination model, while Nandra \etal (1997) showed that the inner regions of BLS1s also appear to be observed nearly face-on. Reverberation results (Kaspi \etal 2000; Peterson \etal 2000) find that the BLRs of NLS1s and BLS1s have comparable sizes, while NLS1s have virial masses that are one order of magnitude smaller than BLS1s. This result shows that the size of the BLR does not scale with the mass of the central black-hole but with the luminosity (Laor 1998) which is connected to the accretion rate. Dewangan \etal (2001a,2001b) suggested that both
the steeper X-ray emission and narrower width of the H$\beta$ line, and also the anti-correlation between the slope of the X-ray spectrum and the width of the H$\beta$ line could be due to the variation in the fractional accretion rate. In this scenario, the higher fractional accretion rate producing an increased accretion disk emission in the soft X-rays and a steeper power law in the hard X-rays, results in a higher radiative pressure per unit gravitating mass. Thus, for a given density distribution of diffuse material, the BLR clouds will form further out in a higher-intensity radiation field, resulting in the narrower emission lines.

Variability studies using long \asca (Advanced Satellite for Cosmology and Astrophysics) observations have been a valuable tool to explore the nature of the soft X-ray excess emission and other spectral components of NLS1s (Turner \etal 2001b; Romano \etal 2002). A 35-day long \asca observation of a NLS1 galaxy, Akn~564, revealed that a slower varying soft excess component is superimposed on a fast varying continuum component (Turner \etal 2001b). Similar results are also inferred from a 12-day \asca observation of another NLS1 galaxy Ton~S180 (Romano \etal 2002).

IRAS~13224-3809 is an extremely variable NLS1 galaxy at  a redshift of 0.06667 and with soft ($0.1-2.4\kev$) X-ray luminosity of $3\times10^{44}\ergsec$ (Boller \etal 1993). The $\FWHM$ of the $\hbeta$ line of this source is only $\sim700\kms$ (Boller \etal 1993; Leighly 1999b) which is comparable to the width of the forbidden line [O~III]$\lambda5007$.  \rosat observations (Boller \etal 1993) showed a complex soft X-ray emission that was very steep ($\Gamma_{\rm X} \sim 4.4$) and rapidly variable (change in intensity by a factor of 2 in $\sim800\s$). Subsequent \asca observations in 1994 confirmed the complex soft X-ray emission and variability (Leighly \etal 1997; Leighly 1999a,1999b). A 30-day \rosat HRI monitoring of IRAS~13224-3809 has revealed the most extreme and multiple giant amplitude X-ray variability (Boller \etal 1997).

In this paper we present the results from a 10-day \asca observation 
of the NLS1 galaxy IRAS~13224-3809. In Sect.~\ref{datared} we describe our 
observations and data reduction. In Sect.~\ref{timevar} we discuss the time 
variability of the source. We analyze the mean spectrum in Sect.~\ref{meansp}  
and time resolved spectra in Sect.~\ref{timeselsp}. In  Sect.~\ref{fluxressp} we 
present flux resolved spectroscopy. In Sect.~\ref{comp} we compare our results for IRAS~13224-3809 with the properties of Akn~564 and Ton~S180. Finally, we discuss the results in Sect.~\ref{discuss} and summarize our results in Sect.~\ref{summary}.

\section{Observation \& Data Reduction  \label{datared}}
\asca consists of four focal plane detectors, two CCDs (the solid-state 
Imaging Spectrometers, SIS0 and SIS1, $0.4-10\kev$, Bruke \etal 1991) 
and two GISs (the Gas Imaging Spectrometers, GIS2 and GIS3, $0.7-10\kev$, 
Ohashi \etal 1996, and references therein). All the four detectors 
operate simultaneously. \asca observed IRAS~13224-3809 (Principal Investigator:  K. M. Leighly) for a total duration of
$\sim834{\rm~ks}$ starting from JD=2451731.562 (for the screened data)
in the 1CCD mode.
The data were reduced using standard techniques (Revision 2). Data screening yielded
an effective exposure time of $228{\rm~ks}$ for SIS0, $226{\rm~ks}$
for SIS1, and $271{\rm~ks}$ for both the GISs. The mean SIS0 count rate
was $(5.063\pm0.061)\times10^{-2}{\rm~count~s^{-1}}$, which is about $30\%$
higher than the SIS0 count rate of $(3.051\pm0.095)\times10^{-2}{\rm~count~s^{-1}}$  found during a previous {\it ASCA} observation
in 1994 (Leighly et. al. 1997; Leighly 1999a, 1999b). 

After the launch in 1993, \asca SIS detectors degraded gradually in efficiency
at lower energies, due to the increased dark current levels and charge transfer 
inefficiency (CTI). This degradation resulted in SIS spectra which diverge from each other and from the GIS data. The instruments can diverge by as much as $40\%$ for energies below $0.6\kev$ for data taken in 
2000 January\footnote{see http://heasarc.gsfc.nasa.gov/docs/asca/watchout.html}. The degradation in efficiency is not well understood and it could not be corrected for by any of the software at the time of writing this paper. There has been a non-linear evolution of the SIS CTI during the last phase of \asca observations (AO-8). The SIS team has revised the calibration of the non-uniform CTI effect and released a new calibration file ($sisph2pi\_130201.fits$) on 2001 March 29. The IRAS~13224-3809 data were calibrated using the above revised calibration file.

The divergence of the SIS detectors at low energies can be compensated for in the spectral analysis by employing the technique of Yaqoob \etal (2000), who provide an empirical correction by parameterizing the efficiency loss as a time-dependent absorption 
term\footnote{see http://lheawww.gsfc.nasa.gov/$\sim$yaqoob/ccd/nhparam.html} (``excess \NH''). The correction for SIS0 follows a linear relationship, $\NH(\mbox{SIS0}) = 3.635857508 \times 10^{-8}\;  (\mbox{
{\tt T}}
-3.0174828 \times 10^{7}) \; 10^{20}$ cm$^{-2}$, where {\tt T} is the
average of start and stop times of the observation measured in seconds
since launch. The SIS1 excess absorption term does not follow the simple linear
relationship with time but it is found that a slightly larger 
absorption column can be applied to the SIS1 data so that both the 
SIS detectors agree well at lower energies. For the observations of 
IRAS~13224-3809, $\NH(\mbox{SIS0}) =7.53 \times
10^{20}$ cm$^{-2}$, where $T=2.37\times10^{8}\s$ and we adopted 
$\NH(\mbox{SIS1}) = 1.05 \times
10^{21}$ cm$^{-2}$.

\section{Time Variability \label{timevar}}
 Light curves were extracted using bin sizes of 500~s
 in the band ($0.7-10{\rm~keV}$) from the SIS data and 5000~s in the
 soft ($0.7-1.3{\rm~keV}$) and hard ($1.3-10{\rm~keV}$) bands from both
 the SIS and GIS data. The soft and hard bands were chosen to have similar
 signal-to-noise and to separate approximately the two spectral
 components -- soft excess and power law (see Sect.~\ref{meansp}). The exposure
 requirements for the light curves were that the bins be at least $50\%$
 and $10\%$ exposed in each instrument for the 500~s and 5000~s curves,
 respectively. Background light curves were  extracted from the source 
 free regions and subtracted from the source light curves after 
 appropriate scaling to compensate for different sizes of extraction regions.
 We combined the 500~s light curves from the two SIS
 detectors only, and 5000~s curves in each band from all the four detectors.
 The observed counts correspond to a mean observed flux of $6.7\times10^{-13}\funit$, and luminosity of $5.9\times10^{42}\ergsec$ (assuming $H_0 = 75\Hunit$, $q_0 = 0.5$) in the $2-10\kev$ band.

  \begin{figure*}
     \centering
     \includegraphics[angle=-90,width=13cm]{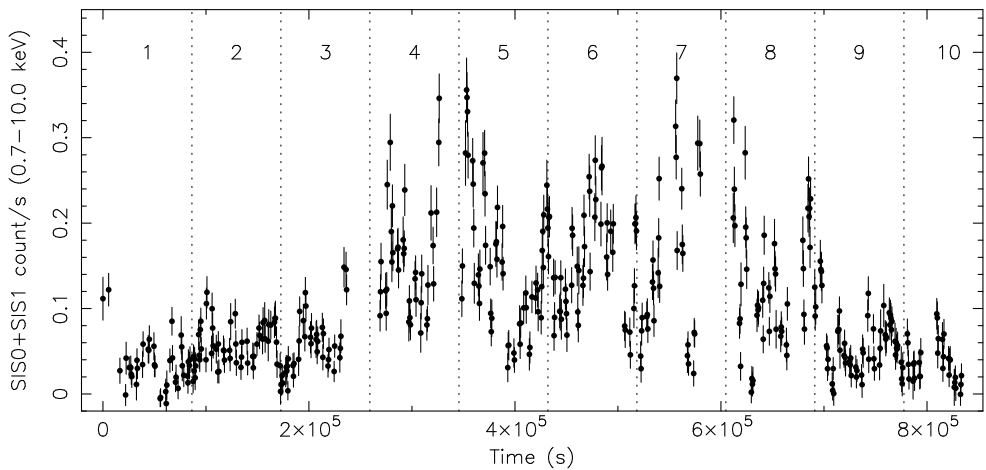}
     \caption{
     SIS0+SIS1 light curve (in 500\s bins) of IRAS~13224-3809 obtained from the \asca data taken during 2000 July 6--15. The light curves were extracted in the energy band of $0.7-10\kev$. The horizontal axis shows the time in seconds from the beginning of the observation (MJD $= 51731.188273$ day).
      }
   \label{2000lc}
\end{figure*}

\begin{figure*}
      \centering
      \includegraphics[angle=-90,width=13cm]{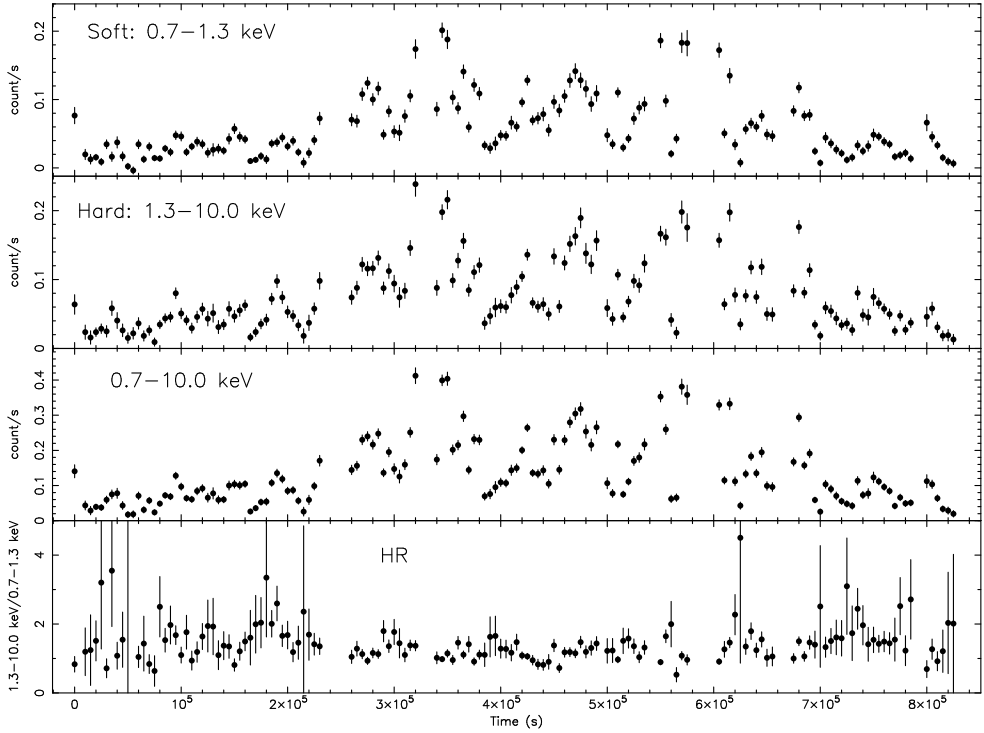}
      \caption{Light curves of IRAS~13224-3809 with $5000\s$ bins. The 
      top panel is the SIS soft-band ($0.7-1.3\kev$) light curve; the second
      panel is the SIS+GIS hard-band ($1.3-10\kev$) light curve; the third 
      panel is the total band ($0.7-10\kev$) light curve. The last panel shows
      the hardness ratio defined as the ratio of the count rates in the 
      $1.3-10\kev$ and $0.7-1.3\kev$ bands. The light curves have been 
      corrected for the background contribution. The horizontal axis shows the
      time in seconds from the beginning of the \asca observation 
      (MJD = $51731.188273$).
      }
      \label{hdr5000}
\end{figure*}

 Fig.~\ref{2000lc}  shows the background subtracted $0.7-10\kev$  band 
 light curve with 500~s bins.
 Fig.~\ref{hdr5000} shows the background subtracted 
 SIS soft-band 
 and SIS+GIS hard-band
 light curves in 5000~s bins. Also shown in Fig.~\ref{hdr5000} is the $0.7-10\kev$ band light
 curve, and hardness ratio (HR) defined as the ratio of count rates in the
 $1.3-10{\rm~keV}$ and $0.7-1.3{\rm~keV}$ bands. 

\begin{figure*}
        \centering
        \includegraphics[angle=-90,width=8cm]{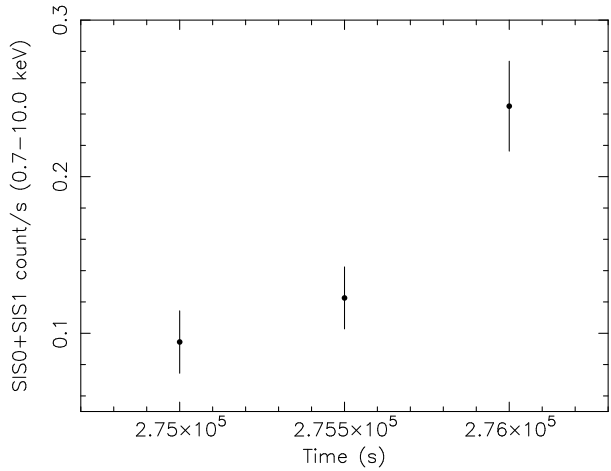}
       \caption{
       The most rapid variability event observed during the 10-day \asca observation of IRAS~13224-3809. The light curve is  an expanded view of the light curve shown in Fig.~\ref{2000lc} during the variability event.}
       \label{most_rapid_event}
\end{figure*}

\begin{figure*}
        \centering
        \includegraphics[angle=-90,width=13cm]{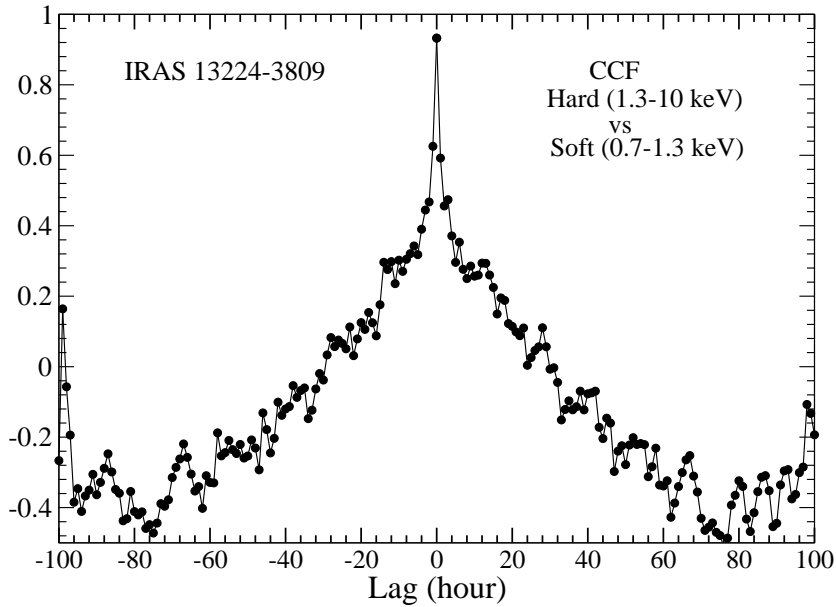}
       \caption{Cross-correlation function of hard ($1.3-10\kev$) flux with respect to the
soft ($0.7-1.3\kev$) flux observed during the year 2000.}
     \label{crosscor}
     \end{figure*}

\begin{figure*}
        \centering
\includegraphics[angle=-90,width=15cm]{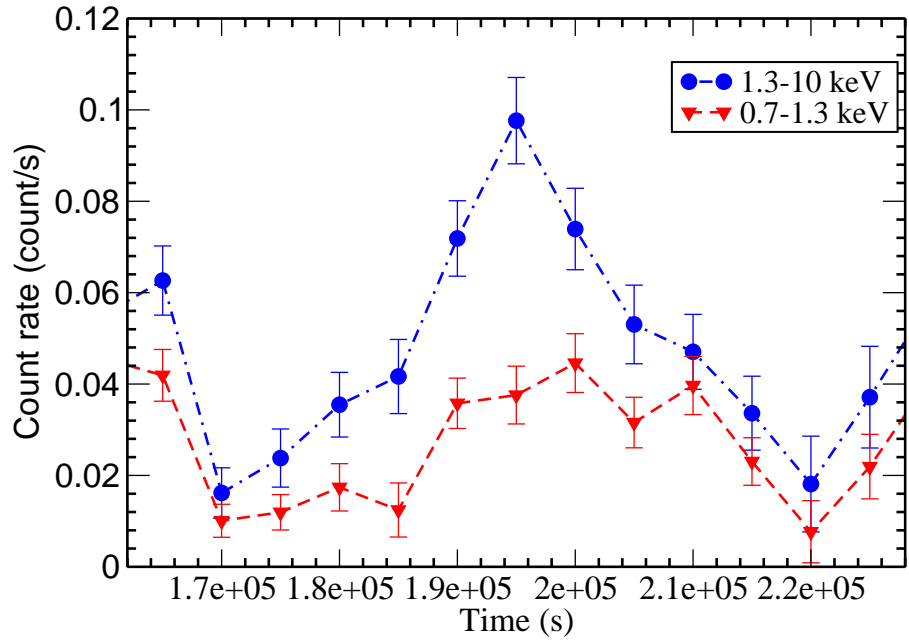}
  \caption{An X-ray flare, in the soft and hard bands, from IRAS~13224-3809 observed $1.9\times10^{5}\s$ after the beginning of the observation in the year 2000.}
\label{lc_event}
\end{figure*}


\begin{figure*}
	\centering
	\includegraphics[angle=-90,width=13cm]{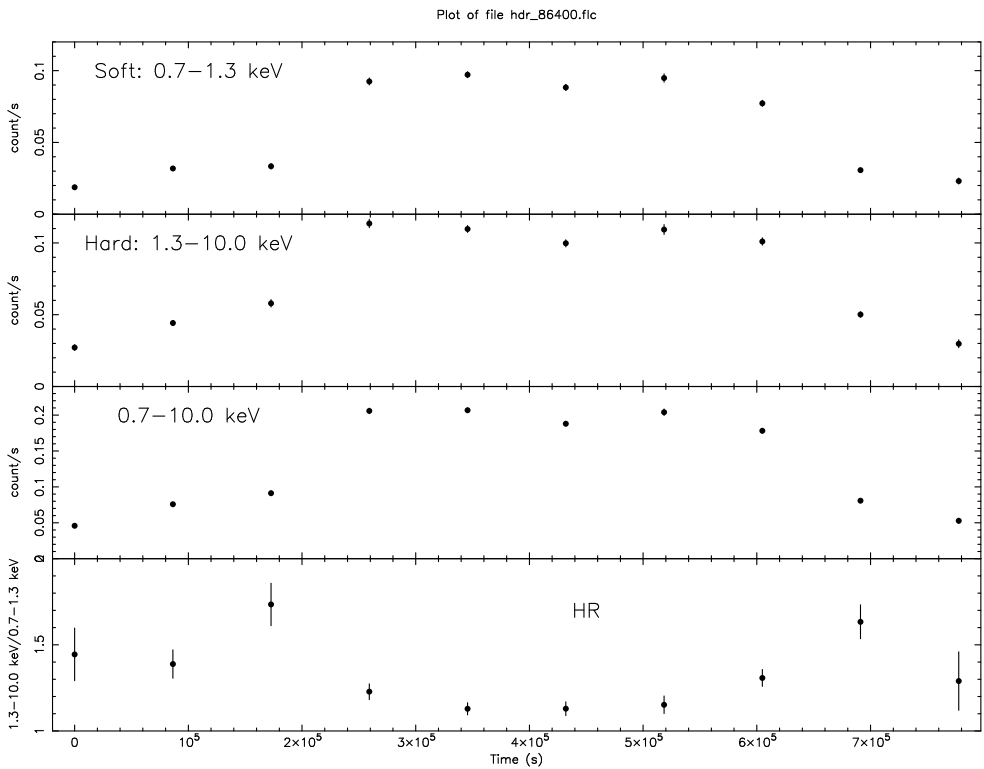}
	\caption{Same as Fig.~\ref{hdr5000} except that the light curves 
	are shown with time bins of 1~day.
	}
	\label{hdr86400}
\end{figure*}	

 The light curves binned to $5000\s$ (Fig.~\ref{hdr5000})
 show trough-to-peak variations in the count rate 
 by a factor of $\ga25$ in the soft band, $\sim 20$ in the hard band. 
 The light curve sampled on $500\s$ (Fig.~\ref{2000lc}) reveals even higher amplitude variations 
 due to fast flickering, with a maximum amplitude of variability of a factor 
 $\ga37$. Close up examination of the light curve in Fig.~\ref{2000lc} reveals several rapid variable 
 events. The most rapid event, shown in Fig.~\ref{most_rapid_event},  
 occurred at $2.755\times10^{5}\s$ after the beginning of the observation. 
 The count rate increased from $0.12$ to $0.24\ctssec$ just within $500\s$. 
 Several variable events with a change in the count rate by a factor of $\sim3$ on a
 timescale of $3000-5000\s$ have been detected e.g., events seen at 
 $3.265\times10^{5}\s$, $3.69\times10^{5}\s$, and $5.775\times10^{5}\s$. 
 The soft and hard X-ray light curves, shown in Fig.~\ref{hdr5000} , show similar variability
 properties. We have calculated the cross correlation function (CCF) of the hard X-ray
 ($1.3-10\kev$) flux with respect to the soft X-ray ($0.7-1.3\kev$) flux. 
 The CCF is plotted in Fig.~\ref{crosscor}, which shows strong correlation between the hard and 
 soft X-ray fluxes without any significant time delay. Since the power-law component 
 contributes $\sim32\%$ of the total flux in the soft ($0.7-1.3\kev$) band as 
 inferred from the mean spectrum (see Sect.~\ref{meansp}), the observed correlation is 
 partly due to variation in the power-law component alone. However, variability 
 of the power-law flux alone is not sufficient to explain the soft X-ray variability. 
 This can be seen from Fig.~\ref{lc_event} which shows an expanded view of the flaring event 
 seen at $1.9\times10^{5}\s$ after the beginning of the observation. 
 The $0.7-1.3\kev$ band flux changed by a factor of $\sim4$ from 
 $0.010\pm0.003{\rm~count~s^{-1}}$ to $0.044\pm0.006{\rm~count~s^{-1}}$, while 
 the $1.3-10\kev$ band flux changed by a factor of 6 from $0.016\pm0.005{\rm~count~s^{-1}}$ to $0.097\pm0.009{\rm~count~s^{-1}}$. If the soft X-ray flux above the hard X-ray 
 power law remains constant, so that the observed soft X-ray variability is entirely due to 
 changes in the power-law flux alone, then a factor of $\sim 12$ change is required in the power-law flux in 
 the $0.7-1.3\kev$ band. This required change is much higher 
 than the factor of 6 observed in the $1.3-10\kev$ flux. Therefore, the soft-excess 
 flux above the hard X-ray power law must also have changed by a factor of $\sim2.6$ either 
 simultaneously or 
 with a short time delay with respect to the hard $0.3-10\kev$ flux. 

From Fig.~\ref{hdr5000} and Fig.~\ref{lc_event}, it is clear that some events have a sharper rise in hard X-rays e.g., the flaring event seen at $1.9\times10^{5}\s$ after the beginning of the observations, with accompanying change in the hardness ratio, while during other flaring events between $3- 6\times10^{5}\s$, there is no change in the hardness ratio. When the light curves are binned on a timescale of a day, another important type of behavior of the source is observed. Fig.~\ref{hdr86400} shows the light curves and HR with time bins of a day. The HR  appears to increase when the flux is rising or falling but settles down to a lower value when the maximum flux is reached. The hardness ratio tells only the relative changes in the soft and hard bands. Any possible change in the spectral shape will be investigated in Sect.~\ref{timeselsp}.

\subsection{Fractional Variability Amplitude}
The fractional variability amplitude $F_{\rm var}$ and its error
$\sigma_{F_{\rm var}}$ are  defined  as
\begin{equation}
F_{\rm var} = \sqrt{\frac{S^2 - \langle \sigma^2_{\rm err} \rangle}{\langle X \rangle^2} } , \;\;\;\;\;\;\;\;\;\;\;
\sigma_{F_{\rm var}} = \frac{1}{F_{\rm var}}
\sqrt{\frac{1}{2N}} \frac{S^2}{\langle X \rangle^2} .
\end{equation}
(Edelson et al. 2001) where $S^2$ is the total variance of the light curve,
$\langle \sigma^2_{\rm err} \rangle$ is the mean squared error, and
$\langle X \rangle$ is the mean count rate.

First we calculated the fractional variability amplitude of the total  band ($0.7-10\kev$) light
curve with 500~s bins to be $F_{\rm var} = 73.5\pm2.6\%$. This quantity measures
deviations relative to the mean, integrated over the entire duration of the observation. We also calculated $F_{\rm var}$ in the soft ($0.7-1.3\kev$) and the hard ($1.3-10\kev$) bands, from the light curves shown in Fig.~\ref{hdr5000} with 5000~s bins. $F_{\rm var}$ thus calculated is $75.3\pm4.6\%$ in the $0.7-1.3{\rm~keV}$ band, and $66.1\pm3.8\%$ in the $1.3-10{\rm~keV}$ band. 

\begin{figure*}
	\centering
	\includegraphics[width=10cm,angle=-90]{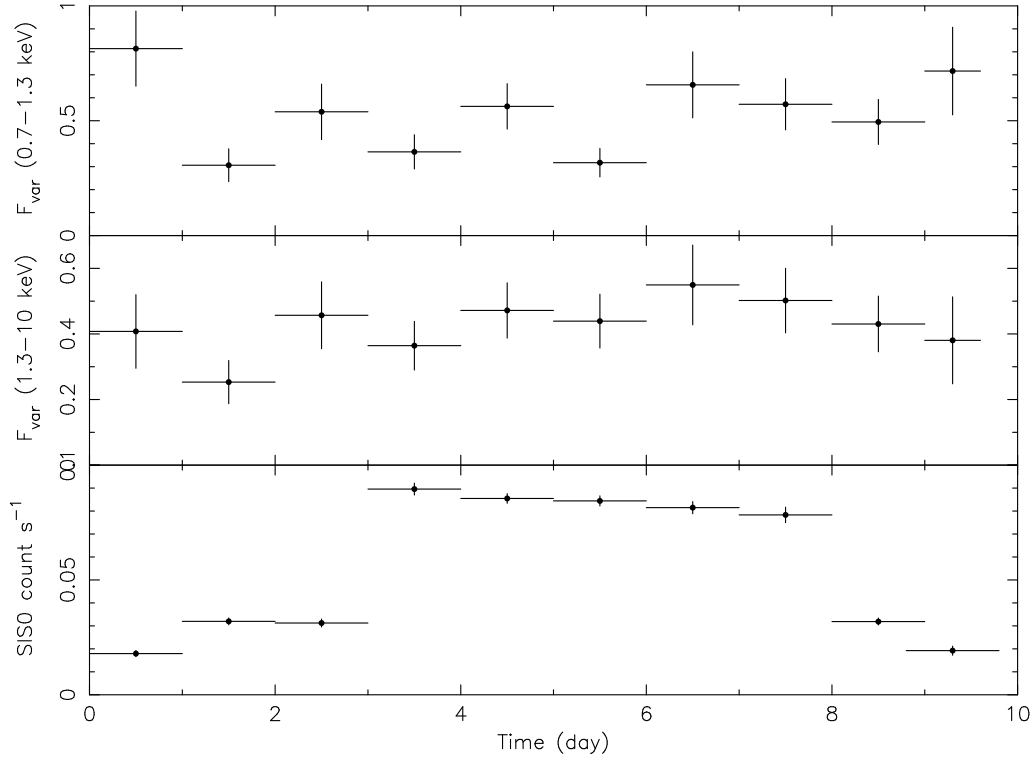}
	\caption{Variability properties of IRAS~13224-3809. 
	First two panels from the top
	show fractional variability amplitude in different energy bands as a 
	function of time. The last panel shows the SIS0 count rate in the 
	$0.7-10\kev$ band.
	}
	\label{fracvar}
\end{figure*}	

 We also measured $F_{\rm var}$ for each day after splitting the light curves with 5000~s bins into 10 evenly-sampled sections across the 10-day \asca observation. Fig.~\ref{fracvar} shows  $F_{\rm var}$ in the soft, and hard bands calculated for each day. 
 The soft-band $F_{\rm var}$  
  changes across the 10-day observation, a constant fit resulting in a minimum $\chi^2$ of 21.88 for 9 dof. 
 The hard-band $F_{\rm var}$, however, does not appear to vary significantly, a constant 
 fit to the hard-band $F_{\rm var}$ curve giving a minimum $\chi^2$ of 9.13 for 
 9 dof. However, there appears to be some similarity in the soft and hard-band $F_{\rm var}$ curves in Fig.~\ref{fracvar}. As already mentioned, the power-law component contributes significantly to the unabsorbed soft-band X-ray flux (see Sect.~\ref{meansp}).  
Therefore, the gross similarity 
 between the soft-band and hard-band $F_{\rm var}$ and a stronger variability of the 
 soft-band $F_{\rm var}$ can be understood if the changes in the soft-band 
 $F_{\rm var}$  are not only caused by variations in the power-law continuum 
 flux but also by the intrinsic variations in the soft-excess component above 
 the power law. It is quite possible that the variations in the above two spectral components are 
 correlated probably with a short time delay.  
 The quantity $F_{\rm var}$ in the soft or hard band, is not correlated 
 with the observed count rate as can be seen in Fig.~\ref{fracvar} suggesting that the variability properties do not depend on the flux level.

\section{The Mean Spectrum \label{meansp}}
For each detector, photon energy spectrum of IRAS~13224-3809
was accumulated from the entire observation. Pulse invariant channels were appropriately grouped for the spectral analysis while considering the degradation in the energy resolution of the SIS detectors. The data from the four instruments were fit simultaneously while keeping the relative normalizations free allowing for the small differences in the calibration of the absolute flux, and differences in the fraction of encircled counts contained in the SIS and GIS extraction cells. The spectral fits were performed with the {\tt XSPEC V11.0.1} package, using response matrices released in 1997 for the GISs, and response files generated using {\tt HEAsoft v5.0.4} for the SISs.

\begin{table*}
\caption{Best-fit model spectral parameters for IRAS~13224-3809 derived from the 10-day \asca observation.}
\label{ana2000}
\begin{tabular}{lcccccccc}
\hline
Data & Model$^1$ & \multicolumn{2}{c}{BB} & \multicolumn{2}{c}{PL} & $f_X~^2$ & $L_X~^3$ & minimum \\
     &  & $kT({\rm~eV})$ & $f_{\rm X}^{\rm BB}~^4$ & $\Gamma_{\rm X}$ & $f_{\rm X}^{\rm PL}~^5$ &  &  & reduced $\chi^2/dof$ \\
       \hline
       Mean & A & -- & -- & $2.02_{-0.08}^{+0.08}$ & 6.7 & -- & -- & 1.331/183 \\
       Mean & B & $130.7_{-3.2}^{+3.4}$ & 5.4 & $2.11_{-0.06}^{+0.05}$ & 6.7 & 16.5 & 15.0 & 1.729/244 \\
       \hline
Day 1 & A & -- & -- & $1.84_{-0.67}^{+0.70}$ & 2.6 & -- & -- & 0.92/36 \\
& B & $130.3_{-22.3}^{+24.1}$ & 1.9 & $1.93_{-0.29}^{+0.38}$ & 2.73 & 6.1
 & 5.6 & 0.84/107 \\
Day 2 & A & -- & -- & $2.09_{-0.32}^{+0.37}$ & 3.8 & -- & -- & 1.04/186 \\
   & B & $125.9_{-12.2}^{+12.3}$ & 3.0 & $2.10_{-0.18}^{+0.19}$ & 3.95 & 9.5
   & 8.8 & 1.03/250 \\
   Day 3 & A & -- & -- & $2.10_{-0.35}^{+0.39}$ &4.0  & -- & -- & 0.91/186 \\
  & B & $114.6_{-18.1}^{+19.1}$ &1.9 & $2.25_{-0.22}^{+0.24}$ & 3.88 & 8.8 & 8.1 & 0.93/250 \\
  Day 4 & A & -- & -- & $2.26_{-0.23}^{+0.24}$ & 8.3  & -- & -- & 1.05/188 \\
   & B &$137.6_{-9.0}^{+9.2}$ & 7.9  & $2.28_{-0.14}^{0.14}$ & 8,61 & 23.5 &
21.9  & 1.14/250  \\
Day 5 & A & -- & -- & $2.43_{-0.20}^{+0.21}$ & 7.4 & -- & -- & 0.93/186 \\
  & B & $130.5_{-7.3}^{7.3}$ & 7.9 &$2.47_{-0.13}^{+0.12}$ & 7.50 & 23.1 & 21.8 & 1.09/250  \\
  Day 6 & A & -- & -- & $2.43_{-0.21}^{+0.21}$  & 7.4  & -- & -- & 1.06/186 \\
  & B & $133.2_{6.9}^{+7.1}$ & 8.9 & $2.31_{-0.12}^{+0.12}$ & 7.82 &17.3  &
   21.9   & 1.09/250 \\
  Day 7 & A & -- & -- & $2.07_{-0.26}^{+0.28}$ &7.5  & -- & -- & 0.95/186 \\
 & B & $135.3_{-9.2}^{+9.6}$  &8.2 & $2.16_{-0.13}^{+0.13}$ & 7.76 & 21.4 & 20.0  & 1.01/250 \\
Day 8 & A & -- & -- & $1.96_{0.20}^{+0.22}$  & 9.3  & -- & -- & 0.98/89 \\
 & B & $134.3_{-9.6}^{+10.2}$ & 6.9 & $2.09_{-0.11}^{+0.13}$ & 9.32 & 22.1
& 20.4  & 1.19/196 \\
Day 9 & A & -- & -- & $1.87_{-0.31}^{+0.31}$ &5.2  & -- & -- &  1.00/53 \\
 & B &$105.7_{-10.9}^{+11.2}$ & 3.2 & $1.90_{-0.17}^{+0.16}$ & 5.26 & 11.1
 & 10.4 & 0.92/154  \\
Day 10 & A & -- & -- & $1.77_{0.61}^{+0.64}$ & 3.5  & -- & -- & 0.89/31 \\
   & B & $98.4_{-16.4}^{+16.7}$ &2.4  &$1.74_{-0.34}^{+0.35}$ & 3.60 & 7.5 &
   7.1  & 1.027/91 \\
\hline
Low$^6$ & A & -- & -- & $1.89_{-0.19}^{+0.17}$ & 3.8 & -- & -- & 1.22/186 \\
    & B & $120.7_{-7.2}^{+6.9}$ & 2.27 & $1.90_{-0.09}^{+0.09}$ & 3.89 & 8.1 & 7.4 & 1.29/250 \\
    Intermediate$^6$ & A & -- & -- & $1.99_{-0.13}^{+0.11}$ & 7.7 & -- & -- & 1.11/186 \\
                 & B & $131.1_{-4.8}^{+4.5}$ & 6.6 & $2.12_{-0.08}^{+0.07}$ & 7.74 & 19.0 & 18.0 & 1.33/250 \\
		 High$^6$ & A & -- & -- & $2.37_{-0.13}^{+0.14}$ & 118.9 & -- & -- & 1.06/186 \\
		      & B & $137.5_{-5.4}^{+5.3}$ & 12.7 & $2.34_{-0.07}^{+0.08}$ & 12.42 & 35.9
		      & 33.6 & 1.24/250 \\
 \hline
 \end{tabular}

$~^1$ Model A is the best-fit redshifted power-law model in the $2-10{\rm~keV}$ band modified by the Galactic absorption. Model B is the combination of redshifted blackbody and power law
       modified by the Galactic absorption. The best-fit parameters for the model B were derived using the total energy band of $0.7-10{\rm~keV}$. \\
$~^2$ Intrinsic flux in the energy band of  $0.7-10{\rm~keV}$ and in the units of $10^{-13}{\rm~erg~cm^{-2}~s^{-1}}$.\\
$~^3$ Intrinsic luminosity in the rest frame and in the energy
band of $0.7-10{\rm~keV}$ and in the units of $10^{42}{\rm~erg~s^{-1}}$.
\\
$~^4$ Intrinsic flux of the soft hump (described by a blackbody) in the $0.7-2{\rm~keV}$ band and in the units of $10^{-13}{\rm~erg~cm^{-2}~s^{-1}}$.\\
$~^5$ Intrinsic flux of the power-law component in the $2-10{\rm~keV}$ band and
in the units of $10^{-13}{\rm~erg~cm^{-2}~s^{-1}}$.\\
$~^6$ The Low, High, and Intermediate states correspond to the observed SIS0 count rates of $\le0.07$, $0.07-0.14$, and $>0.14{\rm~count~s^{-1}}$, respectively. \\
\end{table*}

The spectral shape was first determined by fitting a redshifted power-law model
modified by Galactic absorption (\NH $=4.79\times10^{20}{\rm~cm^{-2}}$; Dickey \& Lockman 1990; Model A) to the data above $2\kev$. An additional absorption term was used for the SIS0 and SIS1 to compensate for the low energy degradation as described in Sect.~\ref{datared}. 
The models for Galactic absorption use the absorption cross-sections of Balu\c{c}inska-Church \& McCammon (1992). For this excercise we used the SIS data in the energy band of $2-7.32\kev$ and GIS data in the $2-10\kev$ band (both in the observer's frame). The power-law
fit yielded $\Gamma_{\rm X} = 2.02\pm0.08$ and a minimum $\chi^2$ of $243.52$ for 183 dof. The errors quoted, here and below, were calculated for the $90\%$ confidence level based on $\chi^2_{\rm min} + 2.71$. 

\begin{figure*}
	\centering
	\includegraphics[angle=-90,width=13cm]{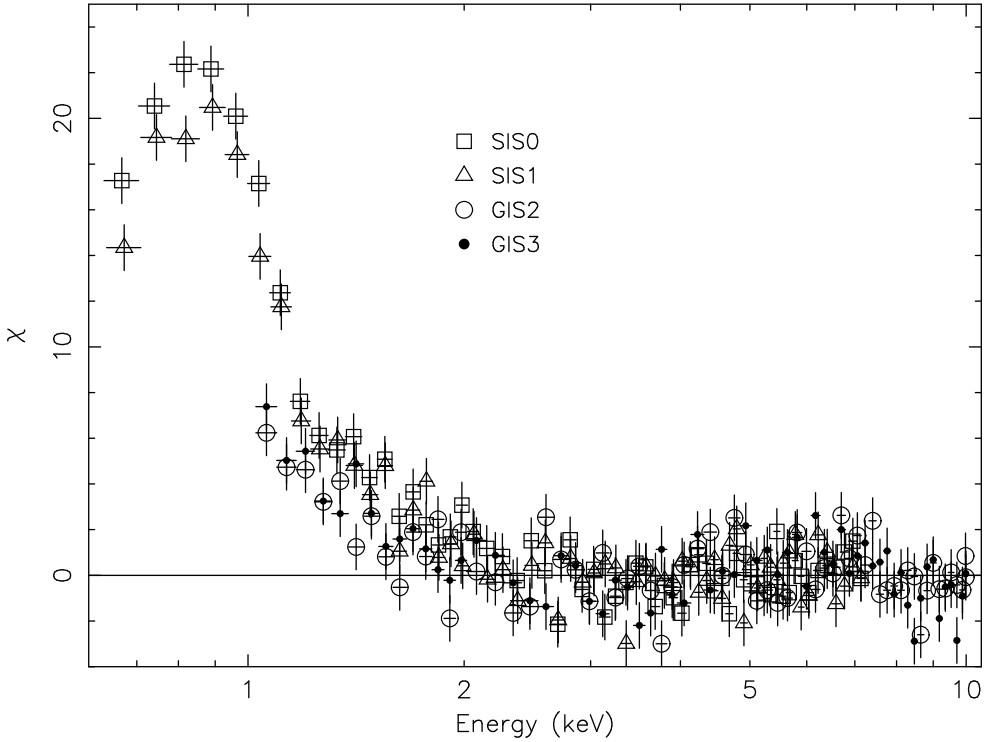}
	\caption{Deviation $\chi$ of the mean spectrum of
	2000 July from the best-fit $2-10{\rm~keV}$ power law
	modified by Galactic absorption where 
	$\chi = \frac{N_{i}^{\rm obs} - N_{i}^{\rm mod}}{\sigma_{i}}$, 
	$N_{i}^{\rm obs}$ is the observed counts in energy channel 
	$i$, $\sigma_i$ is the standard error on  $N_{i}^{\rm obs}$, 
	and  $N_{i}^{\rm mod}$ is the best-fit model counts in channel $i$.  
	}
	\label{mean_dev_pl}
\end{figure*}	

\begin{figure*}
        \centering
	\includegraphics[angle=-90,width=13cm]{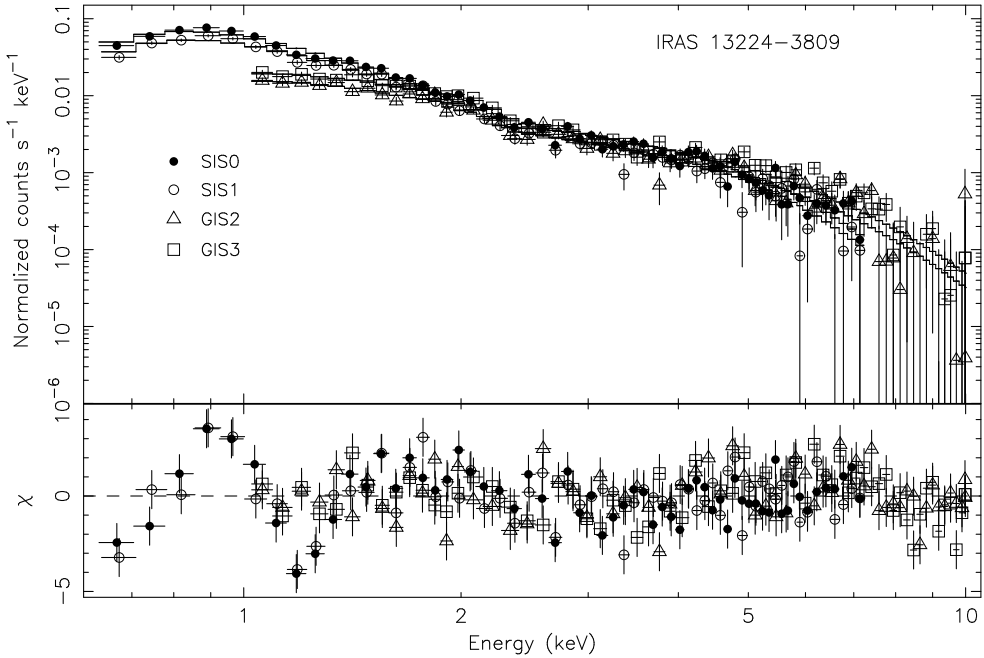}
	\caption{Mean spectra of IRAS~13224-3809 obtained in the year 2000 and the best-fit model -- blackbody and power law modified by the
	        Galactic absorption (Model B) fitted over the entire energy range (top panel) and  deviation of the observed mean spectrum from the 
	best-fit model (bottom panel).
	}
	\label{mean_dev_bb_pl}
\end{figure*}

The results of the fit described above are given in Table~\ref{ana2000} for Model A. We list the best-fitting power-law photon index ($\Gamma_{\rm X}$) in the $2-10\kev$ band, the unabsorbed power-law flux ($f_{\rm X}^{\rm PL}$) in the energy band of $2-10\kev$, the minimum reduced $\chi^2$ ($\chi^2_{\rm red}$), and dof. The deviation of the best-fit model from the observed data is shown in Fig.~\ref{mean_dev_pl} in terms of $\chi = \frac{N_{i}^{\rm obs} - N_{i}^{\rm mod}}{\sigma_{i}}$, where $N_{i}^{\rm obs}$ is the observed counts in energy channel $i$, $\sigma_i$ is the standard error on  $N_{i}^{\rm obs}$, and  $N_{i}^{\rm mod}$ is the best-fit
model counts in channel $i$. The data below $2\kev$ is also shown in Fig.~\ref{mean_dev_pl}. This plot helps to indicate the significant features in the spectrum. A strong soft excess is evident, appearing as a hump of emission below $2\kev$. This feature was also found in the \rosat PSPC observations (Boller \etal 1993), and in the previous \asca observation of 1994 (sequence number 2011000, Leighly 1999b). Hereafter, we refer to this component as the ``soft hump''. Minor calibration problems are also visible mainly below $2\kev$. We do not detect an Fe~K$\alpha$ line from IRAS~13224-3809 in the energy range $6.4-7\kev$ where the deviations are below $3\sigma$ level (see Fig.~\ref{mean_dev_pl}). The $90\%$ confidence upper limit for the equivalent width of  Fe~K$\alpha$ line is found to be $\sim400{\rm~eV}$.

\subsection{The Soft Hump}
We confirm the presence of the soft X-ray excess emission component previously 
observed by Boller \etal (1993) and Leighly (1999b). Similar features are well known in other NLS1 galaxies, e.g. in RE~J1034+393 (Pounds \etal 1995), Akn~564 (Turner \etal 2001b), and Ton~S180 (Romano \etal 2002). In these objects, \chandra LETG results (Ton~S180, Turner \etal 2001a; NGC~4051, Collinge \etal 2001) show that the soft hump is a smooth continuum component, as opposed to a blend of unresolved spectral features. This allows us to choose a continuum component model to parameterize the soft hump emission. We use the blackbody model to characterize the soft hump component as it adequately models the shape and flux of the soft hump. 
We used the SIS data in the range $0.7-7.32\kev$ simultaneously with the GIS data in the range $1-10\kev$, and fitted the redshifted blackbody and power-law model modified by the Galactic absorption (Model B). An additional absorption term as described above was also used for the SIS data. The best-fit blackbody and power-law model yielded a rest-frame temperature $kT = 130.7_{-3.2}^{+3.4} \ev$, absorption corrected blackbody flux, $f_X = 4.34\times10^{-13}\funit$ in the energy band of $0.7-1.3\kev$, and $\Gamma_{\rm X} = 2.11\pm0.05$ for minimum $\chi^2 = 421.98$ for 244 dof. The observed data and the best-fit model are shown in Fig.~\ref{mean_dev_bb_pl}. Also shown in Fig.~\ref{mean_dev_bb_pl} are the deviations of the observed data from the best-fit model. The fit is poor mostly due to calibration problems at low energies and residuals near $1\kev$. The absorption feature near $1\kev$ has been detected in the earlier \asca observation of 1994 and has been interpreted as the blueshifted absorption edges of oxygen (see Leighly \etal 1997). Here we note that the GIS and SIS detectors do not agree at the position of the absorption feature and we do not fit an absorption line or edge model. The uncertainty in the low energy calibration and the degradation in the energy resolution of the SIS detectors make it difficult to explore the absorption feature. We also note that the parameterization of the soft hump as a blackbody does not alter the power-law slope significantly. We find that the power-law continuum contributes $32.3\%$ of the flux in the soft $0.7-1.3\kev$ band in the mean spectrum while the soft hump above the power-law contributes only $2.7\%$ in the $1.3-10\kev$ band. Although the power law has a significant contribution to the soft X-ray flux, the blackbody contribution to the the hard X-ray flux is negligible.

\section{Time Resolved Spectroscopy \label{timeselsp}}

\subsection{Method and Selection Details}
\begin{figure*}
	\centering
	\includegraphics[angle=-90,width=13cm]{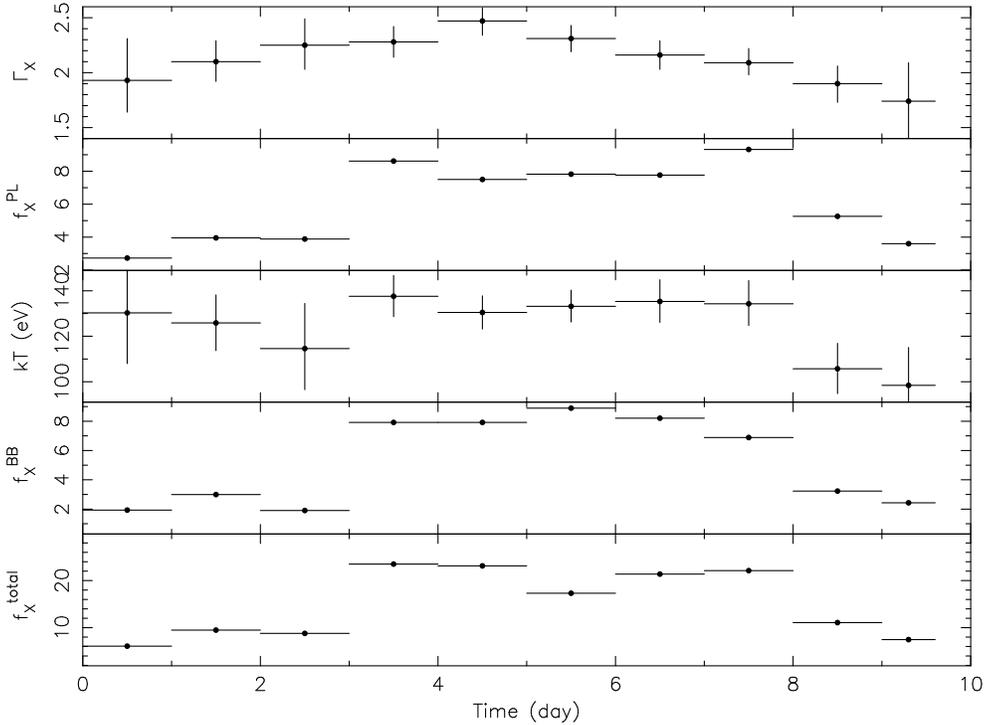}
	\caption{Spectral variability of IRAS~13224-3809. From the top, the time series are the photon index, power-law flux in units of $10^{-13}{\rm~erg~cm^{-2}~s^{-1}}$ and in the $2-10{\rm~keV}$ band, blackbody temperature in eV, blackbody flux in units of $10^{-13}{\rm~erg~cm^{-2}~s^{-1}}$ and in the $0.7-2{\rm~keV}$ band, and the total flux in the $0.7-10{\rm~keV}$ band. All the fluxes have been corrected for the Galactic absorption.
	}
	\label{spec_var}
\end{figure*}

To examine the spectral variations of IRAS~13224-3809, we extracted 10 time-selected spectra across the 9.6-day \asca observation using {\tt Xselect 2.0}. As the source is not bright enough, it was not possible to extract spectra following individual flares and dips with sufficient signal-to-noise ratio. Instead, we chose the sampling timescale of one day, except for the last day of observation for which the sampling time was 0.6~day. The vertical dotted lines in Fig.~\ref{2000lc}  show our 10 intervals within which spectra were extracted. The resulting ``on-source'' average exposure time was $\sim20~{\rm~ks}$ per detector. We set the ancillary response, and response matrix files to be those of the mean spectrum. Again, the energy channels were appropriately grouped to achieve a good signal-to-noise (a minimum of 20 counts per energy channel) while considering the spectral resolution of each detector. All fits were performed by fixing the relative instrument normalizations obtained from the best-fit values from the mean spectrum. The correction for the low energy SIS problem was performed in the same way as for the mean spectrum. All models included the Galactic absorption as before. Fig.~\ref{spec_var} shows the results of our analysis in the form of a time series for the various parameters and are described in detail below. Time assignments of the spectra refer to the mid point of the observation, in days, from the beginning of the observation.

\subsection{Variability of the Continuum}
We fitted each of the 10 time-selected spectra by a redshifted power-law model (Model A) using the $2-7.32\kev$ band for the SIS data and the $2-10\kev$ band for the GIS data (the same data exclusion as for the mean fit). The results of the fits are listed in Table~\ref{ana2000} . In order to explore the time evolution of the soft hump, we also fitted each spectrum by a combination of redshifted blackbody and power-law models (Model B). For these fits, SIS data in the band $0.7-7.32\kev$ and GIS data in the band $1-10\kev$ were used for each time-selected spectrum. The results of these fit are also listed in Table~\ref{ana2000} . As can be seen in Table~\ref{ana2000}, there are no significant changes in the best-fit photon indices obtained from fitting models A and B. Fig.~\ref{spec_var} shows the time series for the photon index, and model power-law flux in the $2-10\kev$ band, blackbody temperature, blackbody flux in the $0.7-2\kev$ band, and total flux in the $0.7-10\kev$ band. Note that the best-fit values for the power-law model plotted in Fig.~\ref{spec_var} are those derived from model B. The best-fit values of $\Gamma_{\rm X}$ range from 1.74 to 2.47 across the 10-day observation, but the day-to-day variations are not significant. A significant variation in the $\Gamma_{\rm X}$ is observed between day~5 and day~9, the change in $\Gamma_{\rm X}$ being $0.57\pm0.21$. We also note that the power-law component dominates the $2-10\kev$ band and the flux variations in this band on a timescale of $\sim$ a day are due to the changes in the continuum level.

\subsection{Variability of the Soft X-ray Hump}
The blackbody flux ($f_{\rm X}^{\rm BB}$) in the $0.7-2\kev$ band roughly follows the power-law flux ($f_{\rm X}^{\rm PL}$) in the $2-10\kev$ band (see Fig.~\ref{spec_var}). The trough-to-peak variation in the soft hump flux is by a factor of $\sim4.7$, while that of the power-law flux is by a factor of $\sim 3.5$ suggesting that the soft hump is more variable than the power-law component on timescales of $\sim$ a week. The blackbody temperature varies from $98.4{\rm~eV}$ to $137.6{\rm~eV}$ across the 10-day observation, but these variations are not significant considering the error bars.
The contribution of the power-law component to the total flux in the $0.7-1.3\kev$
band varies from $24\%$ on day 10 to $48\%$ on day 3 and is not correlated with any of the other spectral parameters.
\begin{figure*}
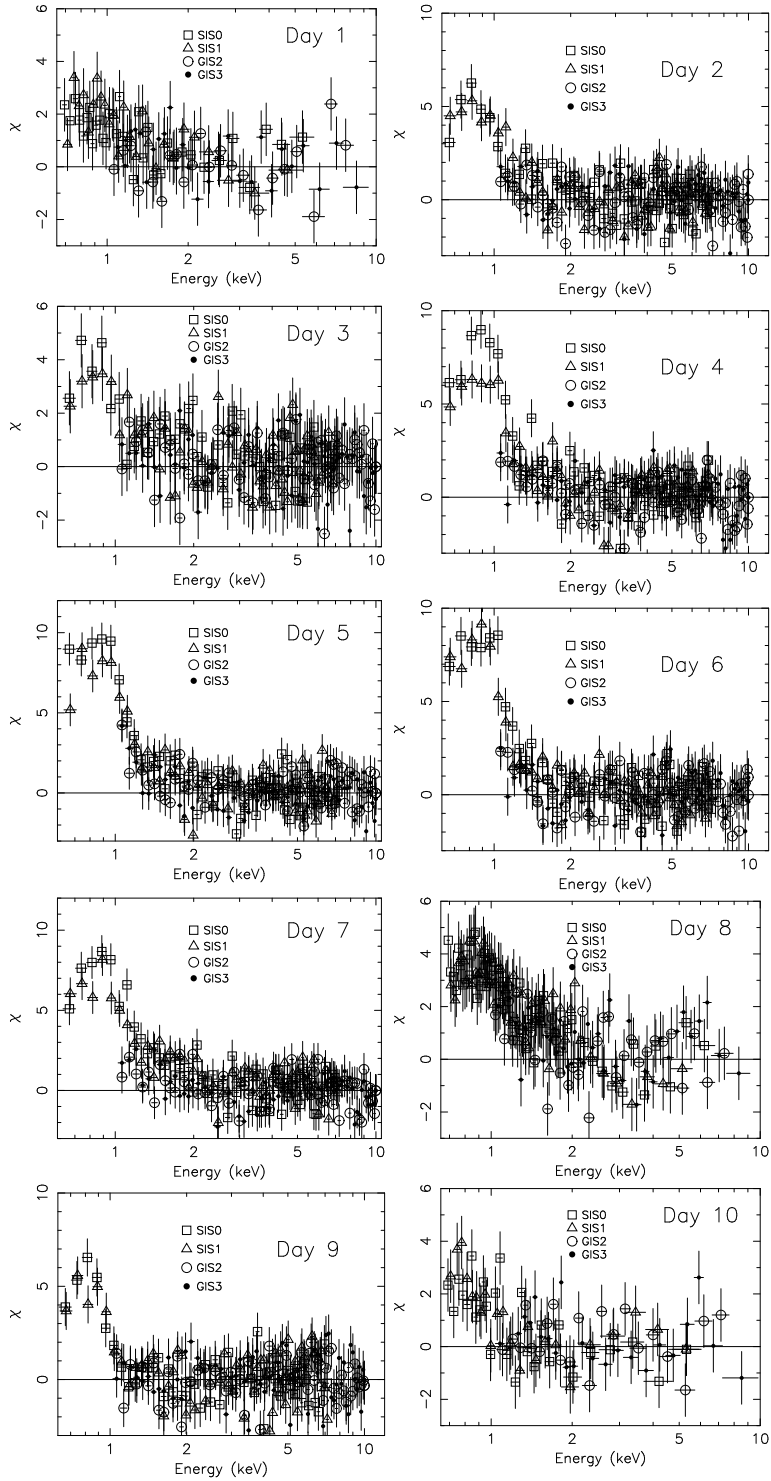

	\centering
	\includegraphics[angle=-90,width=5cm]{day1_relnorm_pl_delchi.ps}
	\includegraphics[angle=-90,width=5cm]{day2_relnorm_pl_delchi.ps}
	\includegraphics[angle=-90,width=5cm]{day3_relnorm_pl_delchi.ps}
	\includegraphics[angle=-90,width=5cm]{day4_relnorm_pl_delchi.ps}
	\includegraphics[angle=-90,width=5cm]{day5_relnorm_pl_delchi.ps}
	\includegraphics[angle=-90,width=5cm]{day6_relnorm_pl_delchi.ps}
	\includegraphics[angle=-90,width=5cm]{day7_relnorm_pl_delchi.ps}
	\includegraphics[angle=-90,width=5cm]{day8_relnorm_pl_delchi.ps}
	\includegraphics[angle=-90,width=5cm]{day9_relnorm_pl_delchi.ps}
	\includegraphics[angle=-90,width=5cm]{day10_relnorm_pl_delchi.ps}
	\caption{Deviations $\chi$ of the time-selected spectra from the 
	$2-10\kev$ best-fit power-law model modified by the Galactic 
	absorption. A strong variation in the soft hump emission above the 
	power law is evident.
	}
	\label{soft_hump_var}
\end{figure*}

\begin{figure*}
        \centering
       \includegraphics[width=10cm]{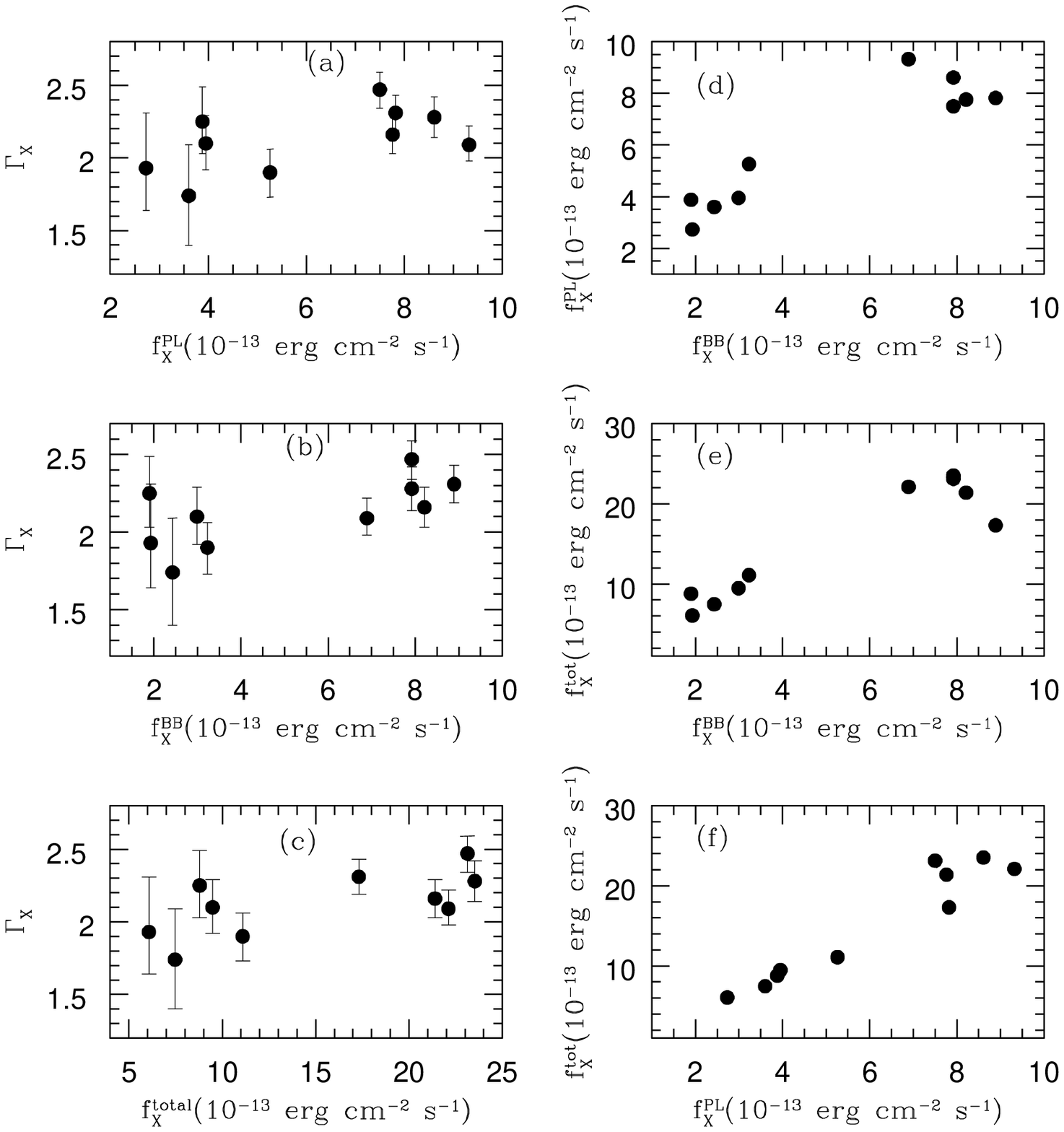}
       \caption{
       Correlations between spectral parameters: (a) $2-10\kev$ photon index ($\Gamma_{\rm X}$) vs $2-10\kev$ power-law flux ($f_{\rm X}^{\rm PL}$), (b) $\Gamma_{\rm X}$ vs blackbody flux in the $0.7-2\kev$ band ($f_{\rm X}^{\rm BB}$), (c) $\Gamma_{\rm X}$ vs total flux (blackbody + power law) in the $0.7-10\kev$ band ($f_{\rm X}^{\rm total}$), (d) $f_{\rm X}^{\rm PL}$ vs $f_{\rm X}^{\rm BB}$, (e) $f_{\rm X}^{\rm total}$ vs $f_{\rm X}^{\rm BB}$, and (f) $f_{\rm X}^{\rm total}$ vs $f_{\rm X}^{\rm PL}$. }
       \label{spec_par_plot}
	\end{figure*}

To examine the time evolution of soft hump flux, we constructed a plot to highlight the variation of the {\it soft hump above the power law}. Fig.~\ref{soft_hump_var} shows the {\it deviations} of the observed data from the best-fit $2-10\kev$ power-law model for all the 10 spectra. The data below $2\kev$ have also been plotted. We note that the soft hump is always evident above the power-law continuum. Strong variations in the soft hump are also evident. 

\subsection{Correlation between spectral parameters}
In order to investigate possible correlation between spectral parameters, we have calculated linear correlation coefficients. Table~\ref{corrmatrix} shows the matrix of linear correlation coefficients and the corresponding significance levels.
\begin{table*}
\caption{Matrix of linear correlation coefficients and the corresponding significance level calculated for the best-fit parameters obtained from the 10 time-selected spectra.}
\begin{tabular}{l|ccccc}
\hline
              & $\Gamma_{\rm X}$ & $f_{\rm X}^{\rm total}$ & $f_{\rm X}^{\rm PL}$ & $kT$ & $f_{\rm X}^{\rm BB}$ \\ \hline
$\Gamma_{\rm X}$    & $1.00(100\%)$ &   $0.67(96.7\%)$ &  $0.57(91.8\%)$ &  $0.65(96.2\%)$ & $0.66(96.6\%)$ \\
$f_{\rm X}^{\rm total}$ & $0.67(96.7\%)$ &   $1.00(100\%)$ & $0.96(99.99\%)$ & $0.68(97.0\%)$ & $0.93(99.97\%)$ \\
$f_{\rm X}^{\rm PL}$    & $0.57((91.8\%)$  &  $0.96(99.99\%)$ &  $1.00(100\%)$ &  $0.63(95.2\%)$ &  $0.92(99.97\%)$ \\ 
$kT$          & $0.65((96.2\%)$  &  $0.68(97.0\%)$ & $0.63(95.2\%)$ &  $1.00(100\%)$ &  $0.70(97.7\%)$ \\
$f_{\rm X}^{\rm BB}$   & $0.66(96.6\%)$ &    $0.93(99.97\%)$ &  $0.92(99.97\%)$ &  $0.70(97.7\%)$ &  $1.00(100\%)$ \\ \hline
\end{tabular}
\label{corrmatrix}
\end{table*}
Fig.~\ref{spec_par_plot} shows plots of spectral parameters.
The most significant correlations are those involving fluxes, both soft hump flux and power-law flux are strongly correlated with the total flux. The power-law photon index is better correlated with the soft hump flux (at $96.6\%$ level) than with the power-law flux (at $91.8\%$ level) suggesting that the shape of the power law is probably more sensitive to the seed photons than to the changes in the power-law flux. The soft hump flux is correlated with the blackbody temperature as is expected for blackbody emission.
\section{Flux Resolved Spectra \label{fluxressp}}
\begin{figure*}
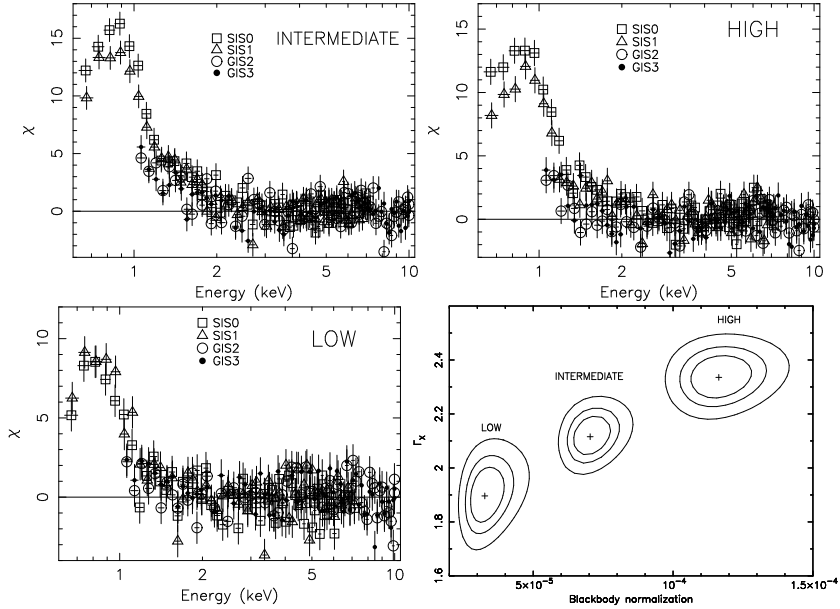

	\centering
	\includegraphics[width=4cm,angle=-90]{2000_intensity_07_14_pl_delchi.ps}
	\includegraphics[width=4cm,angle=-90]{2000_intensity_ge_14_pl_delchi.ps}
	\includegraphics[width=4cm,angle=-90]{2000_intensity_le_07_pl_delchi.ps}
	\includegraphics[width=4cm,angle=-90]{2000_contour_gamma_bbnorm.ps}
	\caption{Deviations $\chi$ of the flux-selected spectra
	from the best-fit $2-10\kev$ power law in the low,
	intermediate, and high state and the confidence
	contours at $68\%$, $90\%$, and $98\%$ levels of the photon indices 
	and blackbody normalizations
	derived from the best-fit Model B.
	}
	\label{2000_flux_res}
\end{figure*}

The soft hump flux and the power-law flux appear to be correlated with the photon index at significance levels of $96.6\%$ and $91.8\%$ (Table 2; see also Fig.~\ref{spec_par_plot}). In order to improve
the statistical significance and to further explore the dependence of spectral parameters, we have carried out spectral analysis at different flux levels. 
We extracted three averaged spectra corresponding to the SIS0 count rates of $\le0.07\ctssec$ (low), $0.07-0.14\ctssec$ (intermediate), and $\ge0.14 \ctssec$ (high). The spectra were analyzed in the same way as before and the results of the spectral fitting are given in Table~\ref{ana2000}. Fig.~\ref{2000_flux_res} shows the deviation of the data from the best-fit $2-10\kev$ power law. In all the three states, the soft hump is present above the power law and is strongly variable. There is a significant change in the power-law slope between the low and the high states, $\Delta \Gamma_{\rm X} = 0.44\pm0.12$. The soft hump flux in the $0.7-2\kev$ band also varied between the two states by a factor of $\sim5.6$, while the power-law flux in the $2-10\kev$ band varied by a factor of $\sim3.2$. Fig.~\ref{2000_flux_res} also shows the contours of allowed values of $\Gamma_{\rm X}$ and blackbody normalization at $68\%$, $90\%$, and $99\%$ confidence levels. It is clear from the contour plot that when the intensity of the source increases, the power law becomes steeper and the soft hump stronger.

\begin{table*}
\caption{Best-fit model spectral parameters for IRAS~13224-3809 derived from the 1994 \asca observation.}
\label{ana1994}
\begin{tabular}{lccccccccc}
\hline
State (SIS0 count rate) & Model$^1$ & \multicolumn{2}{c}{BB} &\multicolumn{2}{c}{PL} & $f_{\rm X}~^2$ & $L_{\rm X}~^3$ & $\chi^2_{\rm red}/dof~^6$ \\
($10^{-2}{\rm~count~s^{-1}}$)   &  & $kT({\rm~eV})$ & $f_{\rm X}^{\rm BB}~^4$ & $\Gamma_{\rm X}$ & $f_{\rm X}^{\rm PL}~^5$ &  &  &
\\
\hline
Mean ($3.5\pm0.09$) &  A & -- & -- & $1.70_{-0.19}^{+0.18}$ & 5.0 & -- & -- & 1.047/183 \\
     & B & $118.3_{-3.7}^{+4.3}$ & 5.5 & $1.71_{-0.12}^{+0.10}$ & 5.2 & 13.0 & 12.5 & 1.186/244 \\
Low ($1.3\pm0.12$) & A & -- & -- & $1.52_{-0.45}^{+0.47}$ & 2.8 & -- & -- & 1.071/186 \\
   & B & $110.1_{-12.2}^{+12.2}$ & 2.1 & $1.56_{-0.20}^{+0.20}$ & 2.8 & 5.8 & 5.6 & 0.974/250 \\
High ($8.8\pm0.31$) & A & -- & -- & $2.45_{-0.28}^{+0.31}$ & 7.5 & -- & -- & 0.973/186 \\
  &  B & $124.0_{-6.1}^{+5.9}$ & 12.7 & $2.20_{-0.15}^{+0.16}$ & 8.0 & 27.5 & 26.9  & 1.173/250 \\
\hline
\end{tabular}

$~^1$ Model A is the best-fit redshifted power-law model in the $2-10{\rm~keV}$ band modified by the Galactic absorption. Model B is the combination of redshifted blackbody and power law
modified by the Galactic absorption. The best-fit parameters for the model B were derived using the total energy band of $0.7-10{\rm~keV}$. \\
$~^2$ Intrinsic flux in the energy band of  $0.7-10{\rm~keV}$ and in the units of $10^{-13}{\rm~erg~cm^{-2}~s^{-1}}$.\\
$~^3$ Intrinsic luminosity in the rest frame and in the energy
       band of $0.7-10{\rm~keV}$ and in the units of $10^{42}{\rm~erg~s^{-1}}$.
\\
$~^4$ Intrinsic flux of the soft hump (described by a blackbody) in the $0.7-2{\rm~keV}$ band and in the units of $10^{-13}{\rm~erg~cm^{-2}~s^{-1}}$.\\
$~^5$ Intrinsic flux of the power-law component in the $2-10{\rm~keV}$ band and
in the units of $10^{-13}{\rm~erg~cm^{-2}~s^{-1}}$.\\
$~^6$ Reduced minimum $\chi^2$/degrees of freedom. \\
\end{table*}

\begin{figure*}
	\centering
	\includegraphics[width=5cm,angle=-90]{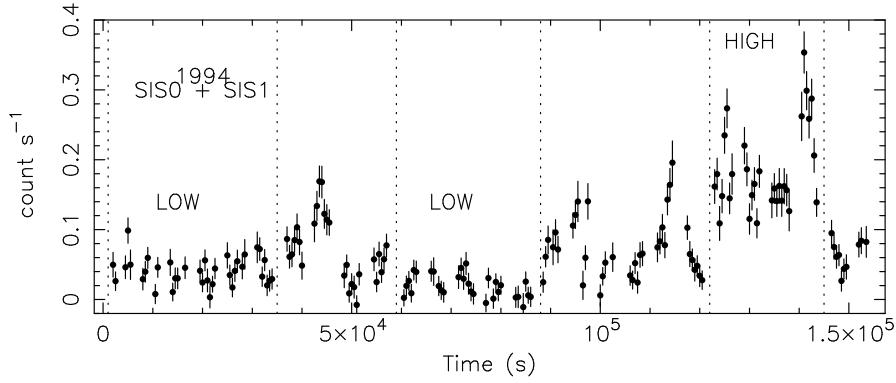} 
	\caption{Light curve of IRAS~13224-3809 sampled with 500~s and 
	in the total energy bands derived from the observation of 
	1994. The vertical dotted lines
	show the time selection for the low and high flux states.
	}
	\label{1994lc}
\end{figure*}

\begin{figure*}
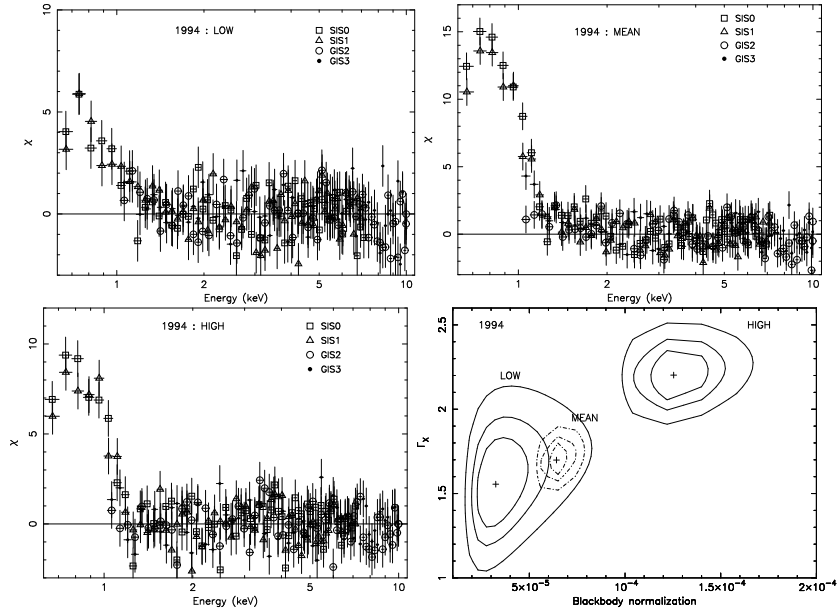

	\centering
	\includegraphics[width=4cm,angle=-90]{1994_intensity_low_pl_delchi.ps}
	\includegraphics[width=4cm,angle=-90]{1994_mean_jfit_pl_delchi.ps}
	\includegraphics[width=4cm,angle=-90]{1994_intensity_high_jfit_pl_delchi.ps}
	\includegraphics[width=4cm,angle=-90]{1994_gamma_bbnorm_cont.ps}
	\caption{Deviation $\chi$ of the spectra, observed in 1994, from the best-fit
	$2-10\kev$ power law. See text and Fig.~\ref{1994lc} for definition of 
	different spectra. Also shown are the confidence contours at 
	$68\%$, $90\%$, and $98\%$ levels of the photon
	index and blackbody normalization from the blackbody and power-law model	fits to the different spectra.
	}
	\label{1994_flux_res}
\end{figure*}

In order to further confirm the above results, we have also analyzed the 
data obtained from an earlier observation in 1994 with \asca. First we 
extracted light curves using bin sizes of 500\s in the total 
band ($0.7-10\kev$) for the SISs. Background light curves were 
also extracted and the source light curves have  been corrected for 
the background contribution. The exposure requirements were that the 
500\s bins be at least $50\%$ exposed. The final light curve was constructed after 
combining the light curves from the two SIS detectors and is shown in 
Fig.~\ref{1994lc}. The timing properties of the source using these data have been 
studied in detail by Leighly (1999a). Our aim here is to show our time 
selection for the low and the high intensity states of the source used for spectral
analysis.
We extracted spectra from the two time intervals where the 
count rate is low (low state) and also from the time interval where the 
source count rate is high (high state). We also extracted spectra from the 
total time interval of the observation (mean spectrum). The spectra were 
analyzed in the same way as before except for the SIS gain 
correction which was not used. The mean, low and high state spectra were fitted by  a redshifted 
power law modified by the Galactic absorption and in the energy band of 
2-10\kev (Model A). The best-fit parameters are listed in Table~\ref{ana1994}. 
The deviations of the observed data from the best-fit power law are shown 
in Fig.~\ref{1994_flux_res}. The soft excess, seen in Fig.~\ref{1994_flux_res}, was again parameterized 
by a redshifted blackbody. The best-fit parameters obtained from the 
blackbody and power law modified by the Galactic observation (model B) 
are also listed in Table~\ref{ana1994}. 
It is evident 
from Table~\ref{ana1994} that the photon index becomes steeper with corresponding 
increase in the blackbody flux in the high state compared to the low state. 
The same result is also seen in the $\chi^2$ contour plots of $\Gamma_{\rm X}$ 
and blackbody normalization in Fig.~\ref{1994_flux_res}. The above result is similar to the results shown in Fig.~\ref{2000_flux_res}.  

\section{Comparison with Akn~564 and Ton~S180 \label{comp}}

A comparison of our results for IRAS~13224-3809 with those obtained for Akn~564 (Turner \etal 2001b) and Ton~S180 (Romano \etal 2002) reveals a broad similarity in the overall shape of the X-ray spectra (strong soft X-ray excess, steep power law) and variability properties. However, there are some important differences: ($i$) An Fe~K$\alpha$ line is not detected from IRAS~13224-3809, while both Akn~564 and Ton~S180 show the presence of an Fe~K$\alpha$ line with a large equivalent width from highly ionized material. ($ii$) The contribution of the power law to the soft hump emission is only $32\%$ in IRAS~13224-3809 while it is $75\%$ for Akn~564 and $72\%$ for Ton~S180. ($iii$) The mean photon index ($\Gamma_{\rm X} = 2.11_{-0.06}^{+0.05}$) of IRAS~13224-3809 is flatter than that of Akn~564 ($\Gamma_{\rm X} = 2.538\pm0.005$) and Ton~S180 ($\Gamma_{\rm X} = 2.44\pm0.02$). ($iv$) The power-law slope appears to be correlated with the soft hump flux in Ton~S180 and IRAS~13224-3809 while the correlation is absent in Akn~564. ($v$) The variability amplitude of the soft hump and power-law components are higher in IRAS~13224-3809 than that of Akn~564 and Ton~S180.

\section{Discussion \label{discuss}}
IRAS~13224-3809 shows the rapid and large amplitude variability seen earlier 
with \rosat (Boller \etal 1993; Boller \etal 1997) and \asca (Leighly 1999a). 
The $0.7-10\kev$ band light curve with $500\s$ bins shows trough-to-peak variation 
by a factor $\ga37$ during our 10-day \asca observation. Rapid X-ray 
variability by a factor of $2-3$ on a timescale of $\sim 2000\s$ has also 
been observed. During the 10-day observation, the light curves sampled with 
$5000\s$ bins show trough-to-peak variations by a factor $\ga25$ in the 
soft band ($0.7-1.3\kev$), and about a factor of 20 in the hard band 
($1.3-10\kev$). Variability events appear to be sharper in the hard 
X-ray band than in the soft X-ray band, and the intensities in the two 
bands are strongly correlated. Changes in the power-law component alone 
are not sufficient to produce the observed correlation but
the soft hump above the power law 
changes its flux simultaneously or with a short time 
delay (see Fig.~\ref{lc_event} and Sect.~\ref{timevar}). 

The mean photon index of the X-ray power law,  
 obtained from the spectral fit to 
the mean spectrum, is $\Gamma_{\rm X} = 2.11_{-0.06}^{+0.05}$. The contribution of the hard X-ray power-law
component to the flux in the soft band ($0.7-1.3\kev$) is only about $32\%$
which is much smaller than that found for Akn~564 ($75\%$, Turner \etal 2001b)
and Ton~S180 ($72\%$, Romano \etal 2002), suggesting that the soft hump
component is more pronounced in IRAS~13224-3809.
Our time resolved spectroscopy reveals variations 
in $\Gamma_{\rm X}$ from $1.74_{-0.34}^{+0.35}$ to $2.47_{-0.13}^{+0.12}$, implying 
a $\Delta\Gamma_{\rm X} = 0.73\pm0.37$. 
The mean power law appears to be 
flatter than that obtained for other NLS1 galaxies like Akn~564 
($\Gamma_{\rm X} = 2.538\pm0.005$, $\Delta \Gamma_{\rm X} = 0.27$, Turner \etal 2001b) 
and Ton~S180 ($\Gamma_{\rm X} = 2.44\pm0.02$, $\Delta \Gamma_{\rm X} = 0.24$, 
Romano \etal 2002) and the variation in $\Gamma_{\rm X}$ is slightly higher in spite 
of the fact that the time span of the \asca observations for the later two 
objects were longer. Thus IRAS~13224-3809 shows higher amplitude variability 
in the power-law slope as well as in the soft hump and the power-law intensity. 
This indicates that the physical parameters governing the X-ray emission 
vary by larger factors in IRAS~13224-3809 compared to that for Akn~564 
and Ton~S180.
\subsection{The $\Gamma_{\rm X}$ - Luminosity Relation}
\begin{figure*}
   \centering
   \includegraphics[width=13cm]{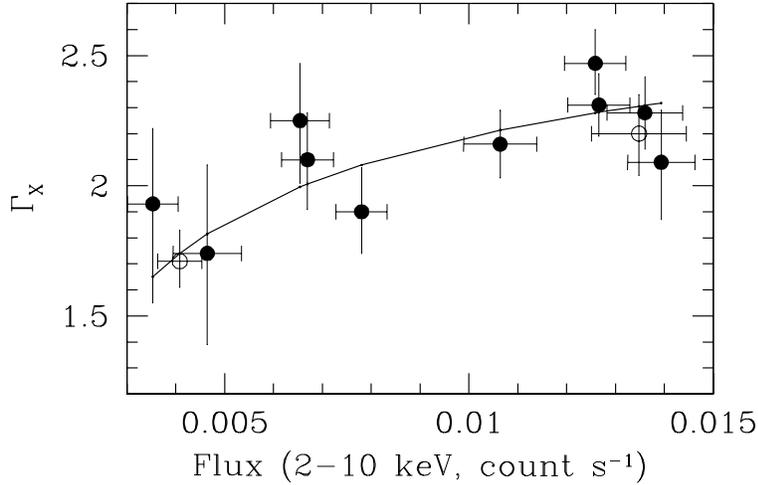}
   \caption{The observed photon index in the $2-10\kev$ band, derived from the time resolved spectroscopy of 2000 (filled circles) and 1994 (open circles) observations, plotted against the observed count rate in the same energy band. The count rate is the average of the GIS2 and GIS3 count rates. Also shown is the best-fit function of the form $\Gamma_{\rm X} = \Gamma_{\rm 0} - K F^{-\delta}$. The best-fit parameters are $\Gamma_{\rm 0} = 3.24\pm0.97$, $\delta = 0.39\pm0.32$. }  
\label{GL}
\end{figure*}

Our time resolved spectroscopy has revealed that the $2-10\kev$ continuum
steepens with increase in the flux. The photon index changes by $\simeq 0.4$ 
with the 
corresponding change in the power-law flux being a factor of $\sim 3.2$ between 
the low and high flux states observed in the year 2000. Similar behavior is seen  
during the 1994 observations. Fig.~\ref{GL} shows 
the plot of $\Gamma_{\rm X}$ against the $2-10\kev$ flux. The steepening of $\Gamma_{\rm X}$ 
with flux has also been observed in a number of Seyfert~1 galaxies (e.g., Singh \etal 1991; Done \etal 2000; Zdziarski \& Grandi 2001; Petrucci \etal 2001; Vaughan \& Edelson 2001, and Nandra 2001) and can be understood in the framework of thundercloud and accretion disk model of Merloni \& Fabian (2001). The basic building blocks of this model are the active regions
above an accretion disk, viewed as magnetic thunderclouds and triggered by magnetic reconnection. The sizes of the active regions are distributed as a power law. Rapid X-ray flares
are produced in the active regions by inverse Compton scattering of soft photons from the disk and $\Gamma_{\rm X}$-luminosity relation of the form $\Gamma_{\rm X} = \Gamma_{\rm 0} - KL^{-\delta}$
is expected (Merloni \& Fabian 2001), where the asymptotic value of $\Gamma_{\rm X}$, $\Gamma_{\rm 0}$, depends mainly on the optical depth $\tau_T$ of the active regions and disk or seed photon intensity. The exponent $\delta$, which determines the amount of spectral variation, is mainly dependent on the spatial distribution of correlated flares. We have fitted the above $\Gamma_{\rm X} - L$ relation to the IRAS~13224-3809 data (see Fig.~\ref{GL}) and obtained the best-fit values $\Gamma_{\rm 0} = 3.24\pm0.97$, $\delta = 0.39\pm0.32$, where the quoted errors are at $1\sigma$ level. Thus for IRAS~13224-3809, the asymptotic photon index is similar within errors to that obtained 
for the Seyfert~1 galaxy MCG-6-30-15 ($\Gamma_{\rm 0} = 2.30_{-0.02}^{+0.63}$; Merloni \& Fabian 2001). Given the large error bars in $\Gamma_{\rm 0}$ for IRAS~13224-3809 as well as for MCG-6-30-15, the coronal optical depth for IRAS~13224-3809 does not appear to be significantly different from that inferred for MCG-6-30-15 ($\tau_T \ge 1.5$).
However, better quality data such as that obtained from monitoring observations with {\it XMM-Newton} are required to make a firm conclusion.
In the framework of thundercloud model a smaller covering fraction of the 
active regions is required for IRAS~13224-3809 than for MCG-6-30-15. The smaller the covering fraction, the larger the observed variability and greater the chance of a large flare to occur (Merloni \& Fabian 2001).

\subsection{The Soft Hump Emission}
Our time resolved spectroscopy has revealed that the soft hump emission and the power-law flux are correlated (see Figs.~\ref{spec_var},~\ref{2000_flux_res},~\ref{1994_flux_res}). Pounds \etal (1995) first noted that the reprocessing of hard X-rays is insufficient to produce the observed soft hump emission of NLS1 galaxies and the soft hump emission could be the intrinsic disk emission resulting from near or super-Eddington accretion rates. In this scenario, sharper variability events are expected in the soft X-rays than in the hard X-rays which is contrary to that observed from IRAS~13224-3809 (see Fig.~\ref{lc_event}). On the other hand, in the framework of reprocessing models, sharper variability events are expected in the hard X-rays than in the soft X-rays. Although a correlation between the soft and hard X-ray flux is expected in both the above scenarios, it is difficult to understand the origin of soft X-ray emission either as the intrinsic disk emission or as the reprocessing of hard X-rays.

We propose yet another mechanism which may be partly responsible for the 
observed soft hump emission from NLS1 galaxies. If the accretion rate is 
super-Eddington, the accretion flow is likely to be dense and optically thick. 
The inner regions of such a disk are supported by the strong radiation pressure. Consequently, the disk puffs up and becomes geometrically thick (see e.g. Collin 2001). The electron density of the accretion disk emitting the big blue bump (BBB) can be written as 
\begin{equation}
n_e \sim 3\times{10}^{16}\ T_{5}^{-1}\ \ M_8^{-1}\ \left({L_{\rm X } \over
L_{\rm Edd}}\right)
\left({R \over 10{R}_{ S}}\right)^{-2}\ {\rm cm}^{-3} \mbox{\ \ .}
\label{eq=nR}
\end{equation}
(Collin 2001), where $T_5$ is the temperature of the disk in units of $10^5{\rm~K}$, $M_8$ is the mass of the black-hole in units of $10^8M\sun$, $R$ is the radial distance from the SMBH, $R_S$ is the Schwarzschild radius, $L_{\rm X}$ and $L_{\rm Edd}$ are the X-ray and Eddington luminosity, respectively. For NLS1 galaxies, $T\sim10^{6}{\rm~K}$, $M\sim 10^6M\sun$, thus $n_e \sim 3\times 10^{12}{\rm~cm^{-3}}$ for $R=10R_{\rm S}$. If the disk thickness $H\ga \frac{1}{n_e \sigma_{\rm T}} \sim 10^{12}{\rm cm} \sim 3R_{\rm S}$, then the disk is optically thick for Thomson scattering. Thus for super Eddington rates, the disk can be optically thick and the high energy photons from the corona incident onto the disk will lose energy by direct Compton scattering and the disk electrons will gain energy. Due to the Coulomb interaction, electrons will quickly thermalize thus increasing the disk temperature and hence increasing the soft hump emission. The soft hump emission of NLS1 galaxies, therefore, may consist of intrinsic disk emission, reprocessing of hard X-ray emission by photoelectric absorption and by direct Compton scattering.

\subsection{Correlation between the power-law flux and the soft hump emission}
The soft-band flux and the power-law flux change either simultaneously or with short time delay (see Fig.~\ref{crosscor}). Our time resolved spectroscopy has revealed that the soft hump flux in the $0.7-2.0\kev$ band changes by larger factor ($\sim 4.7$) than the change (by a factor of $\sim 3.4$) in the power-law flux in the $2-10\kev$ band on a timescale of $\sim$ a week. However, on a timescale of $20000\s$ the 
power-law flux changes by a larger factor ($\sim 6$) than the change in the soft 
hump flux (by a factor of $\sim 2.6$) (See Fig.~\ref{lc_event} and Sect.~\ref{timevar}). The above trend suggests that it is the power-law component that is 
responsible for the most rapid ($\la 1000\s$) variability while the soft 
hump dominates the longer timescale ($\ga$ a week) variability. 
Thus the 500~s variability reported here and 800~s variability of IRAS~13224-3809 
reported in Boller \etal (1993) in the \rosat band could be entirely due 
to changes in the power-law component. However, better signal-to-noise data 
over a broad energy band, for example with {\it XMM-Newton} is 
required to varify the above idea. Boller \etal (1993) rejected the 
standard thin accretion disk model, in spite of the good-fit to the \rosat 
PSPC data, on the ground that the standard disk emission cannot produce 
the observed soft X-ray variability as the shortest timescales (e.g. thermal timescale, sonic timescale) possible for standard thin disks are longer than the observed variability timescale by a factor of $2-3$. This apparent problem can be resolved if the observed rapid variability in the \rosat band is due to the changes in the power-law component alone. If the heating of the corona is by magnetic reconnection (Merloni \& Fabian 2001), the variability timescale could be as short as the coronal dissipation timescale which is given by $\tau(R) \simeq \frac{R}{R_S} \frac{M}{10^6 M\odot}\frac{c}{v_{\rm dis}}$ where $R$ is the size of an active region (see Merloni \& Fabian 2001) and $v_{\rm dis}$ is the dissipation velocity, $\frac{c}{v_{\rm dis}} \ga a~ few$. Thus $\tau(R) \ga a~few \times 100\s$ and rapid variability of the power-law flux can be produced via inverse Compton scattering of disk photons in the active regions. An increased flux of the power-law component would further heat the disk due to increased irradiation. However, detailed time dependent accretion disk-corona models are required in order to understand the correlation between the soft hump emission and power-law flux and the variability amplitudes at different timescales.

\section{Summary \label{summary}}
\begin{enumerate}
\item   On a 10-day baseline, the $0.7-10{\rm~keV}$ band flux
        of IRAS~13224-3809 shows trough-to-peak variation
        by a factor $\ge 37$ when sampled using 500~s bins.
        The hard-band ($1.3-10{\rm~keV}$) and soft-band
        ($0.7-1.3{\rm~keV}$) fluxes present trough-to-peak
        variations by factors $\ge25$ and $\sim20$,
        respectively when sampled using 5000~s bins.
\item	The intensities in the soft and hard bands are strongly correlated 
	and are due to changes in both the power-law flux and soft hump 
	emission simultaneously or perhaps with short time ($\sim {\rm a~few} \times 100\s$) delay.	
\item   The fractional variability amplitude is variable
        in the soft and total bands but not in the hard band.
        None of the variability amplitude is correlated with
        the X-ray flux.
\item   We confirm the presence of a ``soft hump'' above the
        power law at energies below $\le 2{\rm~keV}$. The
        power-law component contributes only $\sim32\%$
        of the flux to the soft hump in the
        $0.7-1.3{\rm~keV}$ band. The soft hump
        component shows flux variations down to timescales
        of 1~day, ranging by a factor of 4.7 in the
        $0.7-2{\rm~keV}$ band. The soft hump flux appears
        to be correlated with the power-law flux.
\item   The mean photon index is $2.11_{-0.06}^{+0.05}$.
        Time resolved spectroscopy reveals significant
        changes in $\Gamma_{\rm X}$, $\Delta \Gamma_{\rm X} = 0.57\pm 0.21$ 
	on a timescale
	of 5~days. However, day-to-day variations are
        not significant. Variations in the power-law flux
        on times of $\sim$1~day are not due to changes in
        the power-law slope.
\item   In our daily sampling, the photon index seems to
        be correlated with the flux of the soft hump, but
	due to poor signal-to-noise of the data, a firm
	conclusion cannot be made.
\item   Flux-selected spectral fits reveal that at higher
        flux levels the power law becomes steeper and the soft
	hump and power-law flux higher. The photon index
	changes by $\Delta \Gamma_{\rm X} \sim 0.4$ while the soft
	hump flux changes by a factor of 5.6 and the power-law flux by a
	factor of 3.2 between the low and high flux states.
\item   The 1994 {\it ASCA} observation also reveals
        increase in the $\Gamma_{\rm X}$ by $\ge0.2$, the soft hump flux
	by a factor of $\sim6$ and the power-law flux by a
	factor of $\sim2.8$ from the low to high flux states.
\item   An Fe~K$\alpha$ line is not detected from
        IRAS~13224-3809. The $90\%$ upper limit to the
	equivalent width of Fe~K$\alpha$ is 400~eV.
\end{enumerate}


\begin{acknowledgements}
GCD is pleased to acknowledge partial supports from the Sarojini Damodaran 
International Fellowship Programme and Kanwal Rekhi Scholarship of the TIFR Endowment Fund. We thank the referee Dr. F. Paerels for the comments on this paper. We are grateful to the \asca team 
for their operation of the satellite. This research has made use of data
obtained from the High Energy Astrophysics Science Archive Center (HEASARC), 
provided by NASA's Goddard Space Flight Center.
\end{acknowledgements}


\begin{thebibliography}{}
\bibitem[2001]{BIF01}
        Ballantyne, D.\ R., Iwasawa, K., \& Fabian, A.\ C.\  2001,
       \mnras, 323, 506

\bibitem[1992]{BM92} 
	Balu\c{c}inska-Church, M., \& McCammon, D., 1992, ApJ, 400, 699.

\bibitem[1993]{Bollerea93} 
	Boller, Th., Tr\"{u}mper, J., Molendi, S., Fink, H., Schaeidt, S., 
	Caulet, A, \& Dennefield, M., 1993, A\&A, 279, 53	

\bibitem[1996]{BBF96}
        Boller, Th., Brandt, W.\ N., \& Fink, H.\ 1996, \aap, 305, 53

\bibitem[1997]{BBFF97}
	 Boller, Th., Brandt, W.\ N., Fabian, A.\ C., \& Fink, H.\ 1997, 
	 \mnras, 289, 393

\bibitem[1992]{BG92} 
	Boroson, T.\ A., \& Green, R.\ F., 1992, ApJS, 80, 109	 

\bibitem[1997]{BME97}
        Brandt, W.\ N., Mathur, S.,  \& Elvis, M. 1997, \mnras,
	285, L25

\bibitem[1999]{Brandtea99}
        Brandt, W.\ N., Boller, T., Fabian, A.\ C., \& Ruszkowski, M.\ 1999,
	\mnras, 303, L53
\bibitem[2001]{collin01}
	Collin, S., 2001, ``Accretion and Emission processes in AGN'', World 
	Scientific, in press.
\bibitem[2001]{Collea01}
        Collinge, M.\ J., Brandt, W.\ N., Kaspi, S., Crenshaw, D.\ M.,
        Elvis, M., Kraemer, S.\ B., Reynolds, C.\ S., Sambruna, R.,  \&
        Wills, B.\ 2001, \apj, 557, 2.

\bibitem[1998]{Comastriea98}
        Comastri, A., et al.\  1998, \aap, 333, 31

\bibitem[2001]{Comastriea01}
        Comastri, A., et al.\ 2001, \aap, 365, 400

\bibitem[2001a]{Dewanganea01a}
	Dewangan, G.\ C., Singh, K.\ P., Jones, L.\ R., McHardy, I.\ M., 
	Mason, K.\ O., \& Newsam, A.\ M. 2001a, \mnras, 325, 1616

\bibitem[2001b]{Dewanganea01b}
	Dewangan, G.\ C., Singh, K.\ P., Chavushyan, V., \& Valdes, J.\ R.,
	2001b, ``The Central Kiloparsec of Starbursts and AGN: The La Palma 
	Connection, ASP
       Conference'', Proceedings Vol. 249. Edited by J. H. Knapen, 
       J. E. Beckman, I.
        Shlosman, and T. J. Mahoney. ISBN: 1-58381-089-7. San Francisco: Astronomical
        Society of the Pacific, 2001, p. 290.

\bibitem[1990]{DickeyL90}
        Dickey, J.\ M., \& Lockman, F.\ M. 1990, \araa, 28, 215

\bibitem[\protect\citename{Done et al. }2000]{do00}
Done C., Madejski G. M., Zycki P. T., 2000, ApJ, 536, 213.

\bibitem[1996]{FH96}
	Forster, K., Halpern, J.\ P., 1996, \apj, 468, 565

\bibitem[1989]{Goodrich89}
        Goodrich, R.\ W.\  1989, \apj, 342, 234
	
\bibitem[1998]{Guainazziea98}
        Guainazzi, M., Piro, L., Capalbi, M., Parmar, A.\ N.,
        Yamaguchi, M., \& Matuoka, M. 1998, \aap, 339, 327

\bibitem[1983]{GFM83}
        Guilbert, P.\ W., Fabian, A.\ C., \& McCray, R. 1983, \apj, 266, 466


\bibitem[2000]{K2000}
        Kaspi, S., Smith, P.\ S., Netzer, H., Maoz, D., Jannuzi, B.\ T., \&
        Giveon, U.\ 2000, \apj, 533, 631

\bibitem[1997]{Laorea97}
        Laor, A., Fiore, F., Elvis, M., Wilkes, B.\ J.,
	\& McDowell, J.\ C.\ 1997, \apj, 477, 93

\bibitem[1998]{Laor98}
	Laor, A., 1998, \apj, 505, L83

\bibitem[1991]{Laor91}
        Laor, A.\ 1991, \apj, 376, 90

\bibitem[1997]{Leighlyea97}
	Leighly, K.\ M., Mushotzky, R.\ F., Nandra, K., \& Forster, K., 1997, 
	\apj, 489, L25

\bibitem[1999a]{Leighly99a}            
        Leighly, K.\ M.\ 1999a, \apjs, 125, 297

\bibitem[1999b]{Leighly99b}
	Leighly, K.\ M.\ 1999b, \apjs, 125, 317

\bibitem[1996]{MPJ96}
        Mason, K.\ O., Puchnarewicz, E.\ M., \& Jones, L.\ R.\ 1996, 
	\mnras, 283, L26

\bibitem[2000]{Mathur00}
        Mathur, S.\ 2000, \mnras, 314, L17

\bibitem[1996]{MFR96}
        Matt, G., Fabian, A.\ C.,  \& Ross, R.\ R.\ 1996,
	\mnras, 278, 1111

\bibitem[2001]{MF01}
	Merloni, A., \& Fabian, A.\ C., 2001, MNRAS, 328, 958.
\bibitem[1998]{MBW98}
	Molthagen, K., Bade, N., \& Wendker, H.\ J., 1998, \aap, 331, 925

\bibitem[1997a]{Nandraea97a}
        Nandra, K., George, I.\ M., Mushotzky, R.\ F., Turner, T.\ J.,
        \& Yaqoob, T.\ 1997a, \apj, 476, 70

\bibitem[1997b]{Nandraea97b}
        Nandra, K., George, I.\ M., Mushotzky, R.\ F., Turner, T.\ J.,
        \& Yaqoob, T.\ 1997b, \apj, 477, 602

\bibitem[2001]{NP01}
	Nandra, K. \& Papadakis, I.\ E., 2001, \apj, 554, 710

\bibitem[\protect\citename{Nandra }2001]{nan01}
	Nandra K., 2001, Advances in Space Research, Volume 28, Issue 2-3, p. 295-306.

\bibitem[1996]{Ohashiea96}
        Ohashi, T., et al.\ 1996, \pasj, 48, 157

\bibitem[1985]{OP85}
        Osterbrock, D.\ E., \&  Pogge, R.\ W. 1985, \apj, 297, 166

\bibitem[2000]{P2000}
        Peterson, B. M. et al. 2000, \apj, 542, 161
\bibitem[\protect\citename{Petrucci et al. }2001]{pet01}
	Petrucci et al., 2001, to appear in Proc. of ``X-ray astronomy 2000'',
	(Palermo, Sep. 2000), Eds. R. Giacconi, L. Stella, S. Serio, ASP Conf. Series, in press.

\bibitem[1995]{PDO95}
        Pounds, K.\ A., Done, C., \& Osborne, J.\ P.\ 1995, \mnras,
	277, L5

\bibitem[1991]{Remillardea91} 
	Remillard, R. A., Grossan, B., Bradt, H. V., Ohashi, T., 
	Hayashida, K., Makino, F., \& Tanaka, Y., 1991, Nature, 350, 589.

\bibitem[2000]{RA00} 
	Rodr\'{i}guez-Ardila, A., Binette, L., Pastariza, M. G., 
	\& Donzelli, C. J., 2000, ApJ, 538, 581

\bibitem[1997] {RP97} Rodr\'{i}guez-Pascual, P. M., Mas-Hesse, J. M., 
	\& Santos-Lle\'{o}, M., 1997, \aap, 327, 72.

\bibitem[2002]{Romanoea02}
	Romano, P., Turner, T.\ J., Mathur, S., George, I.\ M., 
	2002, \apj, 564, 162.

\bibitem[1992]{RFM92}
	Ross, R., Fabian, A.\ C., \& Mineshige, S., 1992, \mnras, 258, 189

\bibitem[1991]{SRV91}
	Singh, K.\ P., Rao, A.\ R., \& Vahia, M.\ N., 1991, \aap, 248, 37. 
\bibitem[1998]{TGN98}        
        Turner, T.\ J., George, I.\ M., \& Nandra, K.\ 1998, \apj,
        508, 648

\bibitem[1999a]{TGN99}        
        Turner, T.\ J., George, I.\ M., \& Netzer, H.\ 1999a, \apj,
        526, 52

\bibitem[1999b]{Turnerea99b}
        Turner, T.\ J., et al.\ 1999b, in Proceedings of the 19th
        Texas Symposium on Relativistic  Astrophysics and Cosmology,
        ed. J. Paul, T. Montmerle, \& E. Aubourg (Saclay: CEA), E441
\bibitem[2001a]{Turnerea01}
       Turner, T.\ J., et al.\  2001a, \apj, 548, L13

\bibitem[2001b]{Akn564I}
        Turner, T.\ J., Romano, P., George, I.\ M., Edelson, R.,
        Collier, S.\ J., Mathur, S., \& Peterson, B.\ M.\ 2001b,
        \apj, 561, 131.

\bibitem[1999a]{Vaughan99a}      
        Vaughan, S., Pounds, K.\ A., Reeves, J., Warwick, R.,
        \& Edelson, R.\ 1999, \mnras, 308, L34

\bibitem[1999b]{Vaughan99b}             
        Vaughan, S., Reeves, J., Warwick, R., \& Edelson, R.\
        1999, \mnras, 309, 113

\bibitem[\protect\citename{Vaughan \& Edelson }2001]{ve01}
	Vaughan S. \& Edelson R., 2001, ApJ, 548, 694.

\bibitem[1998]{WB98}
        Wandel, A., \& Boller, Th.\ 1998, \aap, 331, 884

\bibitem[2000]{Yaqoob}
        Yaqoob, T., et al.\ 2000, \asca{} GOF Calibration Memo,
	ASCA-CAL-00-06-01, v1.0

\bibitem[\protect\citename{Zdziarski \& Grandi }2001]{zg01}
	Zdziarski A. A. \& Grandi P., 2001, ApJ, 551, 186.

\end{thebibliography}
\end{document}